\documentclass[aps,superscriptaddress,twocolumn]{revtex4-1}

\usepackage[pdftex]{graphicx}
\usepackage{dcolumn}
\usepackage{bm}
\usepackage{color}
\usepackage{amsmath}
\usepackage{nicefrac}
\usepackage{amssymb} 

\newcommand{\ff}[1]{{\boldsymbol #1}}
\newcommand{\ca}[1]{{\cal #1}}
\newcommand{\bi}{\begin{itemize}}
\newcommand{\ei}{\end{itemize}}
\newcommand{\be}{\begin{equation}}
\newcommand{\ee}{\end{equation}}
\newcommand{\ba}{\begin{eqnarray}}
\newcommand{\ea}{\end{eqnarray}}
\newcommand{\nab}{\boldsymbol \nabla}
\newcommand{\refe}[1]{(\ref{eq:#1})}
\newcommand{\refeq}[1]{Eq.\ (\ref{eq:#1})}
\newcommand{\labeq}[1]{\label{eq:#1}}
\newcommand{\sothree}{\text{SO}(3)}

\usepackage[pdftex,colorlinks=true,linkcolor=blue,citecolor=blue,filecolor=blue]{hyperref}

\begin{document} 
  
\title{Non-Hamiltonian dynamics of indirectly coupled classical impurity spins}

\author{Simon Michel}

\affiliation{I. Institute of Theoretical Physics, Department of Physics, University of Hamburg, Jungiusstra\ss{}e 9, 20355 Hamburg, Germany}

\author{Michael Potthoff}

\affiliation{I. Institute of Theoretical Physics, Department of Physics, University of Hamburg, Jungiusstra\ss{}e 9, 20355 Hamburg, Germany}

\affiliation{The Hamburg Centre for Ultrafast Imaging, Luruper Chaussee 149, 22761 Hamburg, Germany}

\begin{abstract}
We discuss the emergence of an effective low-energy theory for the real-time dynamics of two classical impurity spins within the framework of a prototypical and purely classical model of indirect magnetic exchange:
Two classical impurity spins are embedded in a host system which consists of a finite number of classical spins localized on the sites of a lattice and interacting via a nearest-neighbor Heisenberg exchange. 
An effective low-energy theory for the slow impurity-spin dynamics is derived for the regime, where the local exchange coupling between impurity and host spins is weak. 
To this end we apply the recently developed adiabatic spin dynamics (ASD) theory. 
Besides the Hamiltonian-like classical spin torques, the ASD additionally accounts for a novel topological spin torque that originates as a holonomy effect in the close-to-adiabatic-dynamics regime.
It is shown that the effective low-energy precession dynamics cannot be derived from an effective Hamilton function and is characterized by a non-vanishing precession frequency even if the initial state deviates only slightly from a ground state.
The effective theory is compared to the fully numerical solution of the equations of motion for the whole system of impurity and host spins to identify the parameter regime where the adiabatic effective theory applies. 
Effective theories beyond the adiabatic approximation must necessarily include dynamic host degrees of freedom and go beyond the idea of a simple indirect magnetic exchange. 
We discuss an example of a generalized constrained spin dynamics which does improve the description  but also fails for certain geometrical setups.
\end{abstract} 

\maketitle 

\section{Introduction}
\label{sec:intro}

The coupling between two magnetic moments can be a so-called direct coupling, such as the usually short-ranged quantum Heisenberg exchange interaction or the long-ranged classical dipole interaction, or an indirect coupling \cite{Mat81,Aue94,NR09}.
All indirect coupling mechanisms, e.g., Anderson's superexchange \cite{And50}, double exchange \cite{Zen51,deG60}, the Ruderman-Kittel-Kasuya-Yosida (RKKY) interaction \cite{RK54,Kas56,Yos57}, or more exotic mechanisms \cite{STP13,SBLK13}, have in common that they are derived perturbatively. 
They represent effective interactions between the magnetic moments or spins, generically of the form $J_{\rm eff} \ff S_{1} \ff S_{2}$, where the effective interaction strength $J_{\rm eff}$ is typically more than an order of magnitude smaller than the typical energy scales of the host, in which the spins $\ff S_{1}$ and $\ff S_{2}$ are embedded.

In the RKKY case, for example, two impurity spins are embedded in an electronic host system, typically a metallic Fermi liquid.
To avoid complications due to Kondo effect \cite{Hew93} and its intertwining with the RKKY coupling 
\cite{Don77,JKW81,LVK05,SGP12,SHPM15}, the impurity spins are represented by {\em classical} spin vectors $\ff S_{1}$ and $\ff S_{2}$. 
If the local exchange coupling $K$ of the impurity spins to the local magnetic moments of the electron system is weak as compared to the energy scales of the host, e.g., the Fermi energy, one may use standard perturbation theory to derive the effective RKKY Hamilton function $H_{\rm eff}(\ff S_{1}, \ff S_{2}) = J_{\rm RKKY} \ff S_{1} \ff S_{2}$.
The RKKY interaction $J_{\rm RKKY} = K^{2} \chi_{12}(\omega=0)$ is given in terms of the nonlocal retarded static (zero-frequency) magnetic susceptibility $\chi_{12}(\omega=0)$, which is an oscillatory and decaying function of the inter-impurity distance. 
The condition $H_{\rm eff}(\ff S_{1}, \ff S_{2}) = \mbox{min}$ then provides us with the impurity-spin ground state configuration. 
Obviously, the derivation of simple effective models is only possible if there is a clear separation of energy scales. 

Effective low-energy exchange couplings, like the RKKY interaction, are also employed to predict the real-time spin dynamics in atomistic spin-dynamics theories \cite{SHNE08}.
This is justified, for instance, if only the impurity spin degrees of freedom are driven out of equilibrium so that one stays in the low-energy sector.
In other words, effective low-energy magnetic couplings also govern the real-time dynamics, if the dynamics of the impurity spins is slow compared to the fast electron dynamics and if only the former are excited initially. 
Generally, this argument exploits that a separation of energy scales obviously translates into a separation of time scales.

A sufficiently weak coupling $K$ not only leads to a separation of energy or time scales but also implies that linear-response theory applies, i.e., in first-order-in-$K$ time-dependent perturbation theory \cite{Sak12,BNF12,SP15}. 
Apart from setups which intrinsically prepare non-equilibrium states \cite{FRZ14,LRP18}, such as transport through nano-structures coupled to leads, linear-response theory predicts that effective impurity-spin couplings are ground-state properties of the host. 
In the RKKY case, it is the {\em ground-state} magnetic susceptibility $\chi_{12}(\omega=0)$ that determines the RKKY effective interaction.
Besides this, full linear-response theory and ground-state response functions also describe other effects, such as Gilbert damping or inertia effects \cite{ON06,BNF12,UMS12,SP15,SRP16a,SRP16b}, but those come at higher order in an expansion in the typical memory time scale and can thus be classified as being of secondary importance.

Closely related to linear-response theory is the idea that the state of the host system, at any instant of time $t$, is the ground state for the given configuration of the classical impurity spins ($\ff S_{1}(t), \ff S_{2}(t)$) at this time.
This is reminiscent of the Born-Oppenheimer approach in molecular dynamics with the nuclei treated classically \cite{MH00}. 
Adiabatic dynamics represents another consequence of the weakness of $K$ and the resulting separation of time scales. 
In case of a host system with a gapped electronic structure, the adiabatic theorem rigorously enforces perfect adiabatic dynamics.
In other cases, it is expected to represent an excellent approximation, which is motivated physically by the idea that the host state should be close to the momentary ground state, if the typical relaxation times of the host dynamics are much shorter then the time scale on which the impurity-spin dynamics takes place.

\begin{figure}[t]
\includegraphics[width=0.95\columnwidth]{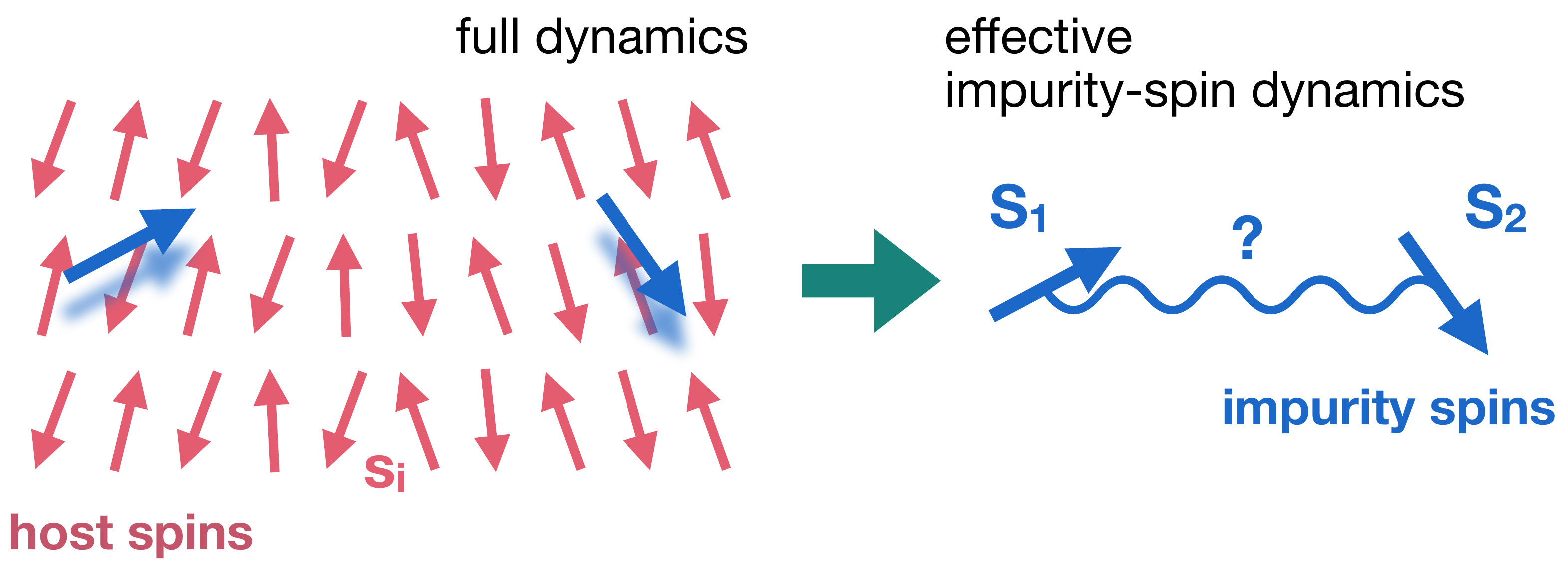}
\caption{
{\em Left:} Two classical impurity spins $\ff S_{1}$ and $\ff S_{2}$ weakly coupled to a system of classical host spins $\ff s_{i}$ localized at the sites $i$ of some lattice. 
{\em Right:} Which type of coupling governs the resulting effective impurity-spin dynamics?
}
\label{fig:problem}
\end{figure}

With the present paper we reconsider this paradigm by studying an even simpler problem: 
We still focus on two classical impurity spins but replace the electronic host system by a system that also consists of classical spins. 
This setup is illustrated in Fig.\ \ref{fig:problem}. 
The host-spin system is given by a classical Heisenberg model with nearest-neighbor interaction $J$ between host spins $\ff s_{i}$ that are localized on the sites $i$ of some lattice. 
In addition, there are two impurity spins coupled to two host spins at sites $i_{1}$ and $i_{2}$ via a local Heisenberg interaction $K$.
We assume a bipartite lattice, such that the $K=0$ host-system ground states are easily found, and we assume a separation of energy scales in the form $|K| \ll |J|$. 

Formally, the {\em static} problem is then treated easily: 
The total energy for a given impurity-spin configuration is $E_{0}(\ff S_{1}, \ff S_{2}) = \mbox{min} \, H(\{\ff s_{i} \}, \ff S_{1}, \ff S_{2})$, where the minimization over all host-spin configurations $\{\ff s_{i} \}$ becomes trivial in the limit $|K| \ll |J|$. 
Therewith we have an effective Hamiltonian $H_{\rm eff}(\ff S_{1}, \ff S_{2}) = E_{0}(\ff S_{1}, \ff S_{2})$ which, for an SO(3) spin-symmetric situation must have the form $H_{\rm eff}(\ff S_{1}, \ff S_{2}) = K f(\ff S_{1} \cdot \ff S_{2})$.
Here, one should note that, opposed to the RKKY setup, our model system is equipped with essentially a single model parameter $K/J$ only, such that in the weak-coupling limit $|K| \ll |J|$ the strength of the effective exchange must scale with $K$, while the scalar (smooth) function $f: \mathbb{R} \to \mathbb{R}$ depends on the system geometry only. 
We will formally derive $H_{\rm eff}(\ff S_{1}, \ff S_{2})$ in the body of the paper. 

Our main interest, however, is focussed on the emergent real-time adiabatic dynamics in case of time-scale separation 
$1/|K| \gg 1/|J|$. 
Assuming that this is fully determined by $H_{\rm eff}(\ff S_{1}, \ff S_{2})$, as the main paradigm suggests, we get the following equations of motion:
$\dot{\ff S}_{r} = \partial H_{\rm eff} / \partial \ff S_{r} \times \ff S_{r}$ for $r=1,2$, i.e., 
\be
\dot{\ff S}_{1} = K f'(\ff S_{1}\ff S_{2}) \, \ff S_{2} \times \ff S_{1} 
\: , \;
\dot{\ff S}_{2} = K f'(\ff S_{1}\ff S_{2}) \, \ff S_{1} \times \ff S_{2} 
\: ,
\label{eq:nn}
\ee
where $f'$ is the derivative of $f$. 
One easily sees that the scalar product $\ff S_{1}\ff S_{2}$ is a constant of motion.
Hence, the impurity-spin dynamics could equivalently be deduced from an effective Hamiltonian of the form $H_{\rm eff}(\ff S_{1}, \ff S_{2}) = J_{\rm eff} \ff S_{1}\ff S_{2}$ with effective interaction $J_{\rm eff} = K f'(\ff S_{1}\ff S_{2})$.
We call this the {\em naive} approach. 

Even in the limit $K\ll J$, the naive approach is shown to fail in many cases, depending on the system geometry.
This is worth mentioning since the approach is very tempting and, furthermore, the reason for its failure is very interesting and instructive:
The pitfall is that the consequences of the assumption that the motion can be described as adiabatic are not taken seriously. 
Assuming that the host system is at time $t$ in its momentary ground state corresponding to the impurity-spin configuration $(\ff S_{1}(t), \ff S_{2}(t))$ means that the state of the whole system (impurity {\em and} host spins) lives in a very small accessible configuration space, parameterized by a product of two Bloch spheres $(\ff S_{1}, \ff S_{2}) \in \ca S_{2} \times \ca S_{2}$. 
This may lead to holonomy effects \cite{Nak98,BMK+03}, i.e., to effects arising from the holonomic constraints responsible for restricting the configuration space. 
Varying the impurity-spin configuration, the ground state of the host-spin system evolves in a geometrically  nontrivial way which mathematically would be expressed in terms of the holonomy of a connection on the manifold of impurity-spin configurations.

In particular, as we have shown in a recent paper \cite{EMP20}, this leads to the appearance of an additional topological spin torque.
The topological spin torque is given in terms of a topological charge density which, when integrated, takes quantized values only.
In Ref.\ \cite{EMP20} we have worked out the general adiabatic spin dynamics (ASD) theory for classical spin systems.
The application of ASD to the case of a single impurity spin ($R=1$), coupled to a host-spin environment and subjected to a local magnetic field has shown that the novel topological spin torque leads to an anomalous precession frequency. 
In the present paper we work out the ASD for the case of $R=2$ impurity spins and analyze the impact of the topological spin torque on the time-dependent indirect exchange.
For the two-spin case one may expect a simple precessional dynamics, similar to Eq.\ (\ref{eq:nn}), but possibly with a renormalized frequency. 
The goal of the present paper is to answer this question, to derive, if possible, the correct effective Hamilton function, and to check the applicability of ASD theory.

The rest of the paper is organized as follows: 
The next section introduces the model, some basic notations and the fundamental equations of motion. 
Sec.\ \ref{sec:asd} briefly reviews the general ASD theory for $R$ impurity spins coupled to a classical host spin system.
The ASD is spin dynamics subject to a formal constraint that enforces adiabaticity. 
We have to carefully specify this constraint. This is done in Sec.\ \ref{sec:hol} and used in Sec.\ \ref{sec:eff} to set up the effective Hamiltonian and to discuss the resulting naive impurity-spin dynamics. 
In Sec.\ \ref{sec:top} we then compute the topological spin torque, and we work out the implications in Sec.\ \ref{sec:dis}. 
The predictions of the ASD, and of the naive approach, can be compared with the numerical solution of the full set of equations of motions. This is done in Sec.\ \ref{sec:res}. 
We discuss the parameter regimes, where the impurity-spin dynamics is close to adiabatic. 
In Sec.\ \ref{sec:bey} we finally discuss an approach which goes beyond the adiabatic approximation and beyond an effective two-spin dynamics. 
Conclusions are given in Sec.\ \ref{sec:con}.

\section{Classical spin model}
\label{sec:mod}

We consider a set of $R$ impurity spins embedded in a lattice of $L$ host spins. 
The host system consists of classical spins $\ff s_{i} = s \ff n_{i}$ of length $s$ and directions given by unit vectors $\ff n_{i} = \ff s_{i} / s$. 
They are localized at the sites $i=1,...,L$ of a $D$-dimensional lattice, and spins $\ff s_{i}$ and $\ff s_{i'}$ interact via an antiferromagnetic Heisenberg exchange $J_{ii'}$. 
The characteristic time scale of the host spin system is set by $1/J$ (we choose units with $\hbar \equiv 1$). 

Impurity spins are given by classical vectors $\ff S_{r}$ for $r=1, ..., R$, and each impurity spin is written in the form $\ff S_{r} = S \ff m_{r}$ with lengths $S = |\ff S_{r}|$ and unit vectors $\ff m_{r} = \ff S_{r}/S$. 
We will focus on the case of $R=2$ impurity spins but develop the theory for the general case of an arbitrary number $R$. 
The impurity spins are locally exchange coupled to the host spins at sites $i_{1}, ..., i_{R}$ of the lattice, and the strength of the local antiferromagnetic (``Kondo'') coupling is denoted as $K$. 

A particular state of this classical spin model is specified by a configuration $(\ff s(t), \ff S(t)) \equiv (\ff s_{1}(t),...,\ff s_{L}(t),\ff S_{1}(t),...,\ff S_{R}(t))$ of host and impurity spins at a time $t$.
Its time evolution is governed by the Hamilton function $H(\ff s, \ff S)$:
\be
H 
=
\frac{1}{2} \sum_{i,i'=1}^{L} J_{ii'} \ff s_{i} \ff s_{i'}
+
K \sum_{r=1}^{R} \ff s_{i_{r}} \ff S_{r}
-
\sum_{r=1}^{R} \ff S_{r} \ff B_{r}
\: .
\label{eq:ham}
\ee
Generically, we have $J_{ii'} = J_{i'i} = J$ between nearest neighbors $i$ and $i'$ only. 
We have also added a term describing a local magnetic field $\ff B_{r}$ coupling to the impurity spin $\ff S_{r}$. 
In most cases, however, we set $\ff B_{r}=0$.
The Hamilton function leads to the following coupled set of non-linear ordinary differential equations of motion, 
\be
  \dot{\ff s}_{i} = \frac{\partial H}{\partial \ff s_{i}} \times \ff s_{i}
  \: , \quad 
  \dot{\ff S}_{r} = \frac{\partial H}{\partial \ff S_{r}} \times \ff S_{r}
  \; , 
\label{eq:hameq}  
\ee
which determine the time evolution of an arbitrary given initial spin configuration.
Note that the lengths of $\ff s_{i}$ and of $\ff S_{r}$ are conserved.
This allows us to absorb constants, like gyromagnetic ratios, in $s_{i}$ and $S_{r}$.
The model (\ref{eq:ham}) can be seen as the classical isotropic multi-impurity Kondo-necklace model
\cite{Don77} or simply as a classical Heisenberg model with a special multi-impurity geometry. 
There is no direct coupling of the impurity spins but an indirect coupling is mediated via the host.

We will study the model in the limit of weak local coupling $K \ll J$. 
In this parameter regime the system exhibits two very different time scales, $K^{-1}$ and $J^{-1}$, such that the fast host spins almost instantaneously follow the motion of the slow impurity spins.
In this adiabatic limit, one can expect a strong conceptual simplification, providing us with an effective theory for the slow degrees of freedom only.

\section{Adiabatic spin dynamics theory}
\label{sec:asd}

Using the notation $\ff n(t) \equiv (..., \ff n_{i}(t), ...)$ and $\ff m(t) \equiv (..., \ff m_{r}(t), ...)$ to characterize the configurations of the fast and of the slow spins at a time $t$ by the respective unit vectors, one can state that the time evolution is adiabatic, if, at any instant of time $t$, the configuration of the fast spins $\ff n(t)$ is the ground-state configuration for the present configuration $\ff m(t)$ of the slow spins at time $t$:
\be
   \ff n(t) = \ff n_{0}(\ff m(t)) \: .
\label{eq:con}
\ee
We expect adiabatic spin dynamics to be realized for $K \ll J$ (and $B_{r} \ll J$).
The condition $\ff n  =  \ff n_{0}(\ff m)$ specifies a hyper surface $\{ \ff n  =  \ff n_{0}(\ff m) \: | \: \ff m \: \mbox{arbitrary} \}$ in $\ff n$-space, see \refeq{con}, i.e., in the product of Bloch spheres, $\ff n \in \prod_{i=1}^{L} \mathbb{S}^{2}$. 
Upon approaching the weak-coupling limit in parameter space, the fast-spin configuration will be more and 
more constrained to this hyper surface.
This means that a strongly simplified description, i.e., adiabatic spin dynamics (ASD), should be possible in this limit.

Using the constraint (\ref{eq:con}), one can define an effective Hamilton function, 
\be
H(\ff s, \ff S) \mapsto H(s \ff n_{0}(\ff m),S \ff m) \equiv H_{\rm eff}(\ff m)
\: , 
\ee
which depends on the slow-spin degrees of freedom only. 
It is therefore tempting to derive the slow-spin dynamics solely from this effective Hamiltonian via the $R$ remaining differential equations
\be
S \dot{\ff m}_{r} = \frac{\partial H_{\rm eff}(\ff m) } {\partial \ff m_{r} } \times \ff m_{r}
\label{eq:na}
\ee
for $\ff m(t)$, while $\ff n(t)$ can be obtained from Eq.\ (\ref{eq:con}).
This constitutes an approach which we will refer to as the {\em naive} theory of adiabatic dynamics.

We have recently shown that the naive approach may lead to incorrect results \cite{EMP20}.
To eliminate the fast spin degrees of freedom correctly, one must rather switch to a Lagrangian formulation and employ the general action principle, $\delta  \int dt \, L (\ff n, \dot{\ff n}, \ff m, \dot{\ff m}) = 0$. 
In this framework one may safely make use the holonomic constraints (\ref{eq:con}) to eliminate the host degrees of freedom and to set up an effective Lagrangian, $L_{\rm eff}(\ff m,\dot{\ff m}) \equiv L(\ff n_{0}(\ff m),(d/dt)\ff n_{0}(\ff m),\ff m,\dot{\ff m})$, for the slow-spin degrees of freedom only.
In terms of the effective Lagrangian, the action principle reads $\delta \int dt \, L_{\rm eff} = 0$, where $\delta$ is variation of the slow-spin configuration $\ff m$ only. 
This provides us with the ASD equations of motion for the slow spins $\ff m_{j}$:
\be
S \dot{\ff m}_{r} = \frac{\partial H_{\rm eff}(\ff m)}{\partial \ff m_{r}} \times \ff m_{r} + \ff T_{r} \times \ff m_{r} \, . 
\label{eq:asd}
\ee
As compared to the ``naive'' adiabatic theory, Eq.\ (\ref{eq:na}), there is an additional term due to a field
\be
  T_{r\mu} =  T_{r\mu}(\ff m,\dot{\ff m}) = \sum_{s\nu} \Omega_{r\mu,s\nu}(\ff m) \dot{m}_{s\nu}  \: ,
\label{eq:tj}  
\ee
where $s=1, ..., R$ and $\mu,\nu \in \{x,y,z\}$. 
Here,
\be
\Omega_{r\mu,s\nu}(\ff m) = 4\pi \sum_{i} s e^{(i)}_{r\mu,s\nu}(\ff m) \: ,
\labeq{omdef}
\ee 
with
\be
e^{(i)}_{r\mu,s\nu}(\ff m)
=
\frac{1}{4\pi}
\frac{\partial \ff n_{0,i}(\ff m) }{\partial m_{r\mu}}
\times
\frac{\partial \ff n_{0,i}(\ff m) }{\partial m_{s\nu}}
\cdot
\ff n_{0,i}(\ff m) 
\: 
\label{eq:cd}
\ee
is a rank-2 tensor for each pair of impurities $r,s$. 
It  describes certain topological properties of the ground state of the fast-spin subsystem on the hyper surface specified by Eq.\ (\ref{eq:con}). 
In fact, each tensor element for fixed $r,s$ defines a topological charge density, which becomes a quantized homotopy invariant, namely a topological winding number $e^{(i)}_{rs}$ with a quantized value $e^{(i)}_{rs} \in \mathbb{Z}$ when integrated.
There is a close analogy to the concept of the skyrmion density \cite{Bra12,FBT+16,SHP+17}. 
Here, however, the skyrmions live on a product of Bloch spheres rather than in Euclidean space. 
Note that the resulting topological spin torque $\ff T_{r} \times \ff m_{r}$ in Eq.\ (\ref{eq:asd}) involves the time derivative $\dot{\ff m}_{s}$. 
It nevertheless respects total-energy conservation, unlike a Gilbert damping term \cite{Gil55,Gil04}.
Details of the derivation of Eq.\ (\ref{eq:asd}) and its interpretation are given in Ref.\ \onlinecite{EMP20}.

\section{Holonomic constraint}
\label{sec:hol}

For the computation of the topological spin torque, the topological charge density (\ref{eq:cd}) is required. 
This is a ground-state property of the host system. 
Similar to a two-point response function, it depends on two fixed positions $i_{r}$ and $i_{s}$, i.e., on the two sites at which the $r$-th and the $s$-th impurity spin couple to the host. 
For a concrete calculation, we need the explicit form of the constraint (\ref{eq:con}). 
This means to find the ground-state configuration of the host spins for an arbitrarily given configuration $\ff m=(\ff m_{1}, ..., \ff m_{R})$ of all impurity spins. 

We start by considering the ground state at $K=0$.
In this case, the host subsystem is described by a classical Heisenberg Hamiltonian $H_{\rm host}(\ff n) = \frac12 s^{2} \sum_{i,i'} J_{ii'} \ff n_{i} \ff n_{i'}$, see \refeq{ham}.
For any choice of the matrix of coupling constants $J_{ii'}$, the Hamiltonian is invariant under SO(3) spin rotations such that the ground-state manifold is highly degenerate. 
We pick an arbitrary ground-state configuration $\ff n_{0} = (\ff n_{0,1},...,\ff n_{0,L})$ which minimizes $H_{\rm host}(\ff n)$, i.e., $H_{\rm host}(\ff n_{0}) = E_{0} = \text{min.}$, where $E_{0}$ is the ground-state energy. 
The SO(3) symmetry implies that $R \ff n_{0} \equiv (R \ff n_{0,1},...,R \ff n_{0,L})$ for any rotation matrix $R \in \sothree$ is a ground state as well: $H_{\rm host}(R\ff n_{0}) = E_{0}$.

At finite $K$, the degeneracy of the ground-state energy is lifted. 
Depending on the strength of $K$, on the given impurity spin configuration $\ff m$, and on the coupling constants $J_{ii'}$, the host ground state can strongly differ from $R\ff n_{0}$ and must be determined numerically in general. 
For weak Kondo coupling $K \ll J$, however, the impurity spins basically act as infinitesimally weak external local magnetic fields, which merely break the host SO(3) invariance and typically favor exactly one state out of the $K=0$ ground-state manifold, without further disturbing the spin configuration of that state.
At the same time, $K \ll J$ just specifies a limit where the ASD is expected to apply -- as will be discussed later by comparing with results from the numerical solution of the full set of equations of motion given by Eq\ (\ref{eq:hameq}).
Hence, we will focus on the weak-coupling limit. 

The host ground state for $K\ne 0$ and $K \ll J$ is obtained by minimization of the Hamilton function $H(\ff n, \ff m) \equiv H(s \ff n, S \ff m)$, where $\ff n_{i} = R \ff n_{0,i}$, for given fixed $\ff m$, and with respect to all $R \in \sothree$. 
Note that this minimization problem is much simpler as compared to a high-dimensional minimization with respect to arbitrary host-spin configurations, which would be necessary beyond the weak-coupling limit.
For the minimization, we can also disregard the magnetic field term in \refeq{ham} as this is independent of $\ff n$ and thus of $R$.
Furthermore, due to the invariance of the inner product $\ff s_{i} \cdot \ff s_{i'}$ under spin rotations, also the Heisenberg term in $H$ is constant. 
Hence, it is sufficient to focus on the Kondo term only:
\be
K \sum_{r=1}^{R}  \, (R\ff n_{0,i_{r}}) \cdot \ff m_{r} \stackrel{!}{=} \mbox{min.}
\: .
\label{eq:mini}
\ee

We make use of the fact that $R \in \sothree$ has the form $R=R_{\ff a}(\varphi)=\exp(\varphi \ff L \ff a)$, where $\ff L$ with components $L_{\alpha} \in \text{so}(3)$ are the real and antisymmetric generators of SO(3), and where the unit vector $\ff a$ specifies the rotation axis and $\varphi$ the rotation angle.
Let us now assume that $\ff n_{0}$ is the desired ground-state configuration for given $\ff m$. 
This implies that the Hamilton function reaches its minimum at $\varphi=0$ for any rotation axis $\ff a$.
With the general relation
\be
\frac{\partial}{\partial \varphi} \exp(\varphi \ff L \ff a) \Big|_{\varphi=0} ( \cdot ) = \ff a \times ( \cdot )
\:, 
\ee
we thus find the following necessary condition for the minimum:
\be
\sum_{r=1}^{R} \ff a \times \ff n_{0,i_{r}} \cdot \ff m_{r} = 0
\: .
\ee
Since this must be satisfied for rotations around an arbitrary axis $\ff a$, we get 
\be
\sum_{r=1}^{R} \ff n_{0,i_{r}} \times \ff m_{r}  = 0
\: .
\labeq{equi}
\ee
This means that the total torque on the impurity spins must vanish for the ground-state configuration $\ff n_{0}$.

We now specialize to the case of host spins on a bipartite lattice with nearest-neighbor interactions $J_{ii'}$, where the ground-state spin structure is collinear, i.e., we have 
\be
\ff n_{0,i}  = z_{i} \ff \eta
\label{eq:coll}
\ee
with $z_{i} \in \{+1,-1\}$ and with a unit vector $\ff \eta$ to be determined from \refeq{equi}.
The host spin structure is invariant under the simultaneous transformation $z_{i} \mapsto -z_{i}$ and $\ff \eta \mapsto - \ff \eta$.
Hence, we can choose the overall sign of $(z_{1},..., z_{L})$ as is convenient, e.g., such that $z_{i_{1}} = +1$.
Physical properties do not depend on this choice.
To fix $\ff \eta$, we insert \refeq{coll} in the equilibrium condition (\ref{eq:equi}). 
This yields
\be
\ff \eta \times \sum_{r=1}^{R} z_{i_{r}} \ff m_{r} = 0
\: .
\labeq{sing}
\ee
We define the (staggered) sum of the impurity-spin unit vectors $\ff m_{r}$
\be
\ff  m_{0} = \sum_{r} z_{i_{r}} \ff m_{r}
\ee
and $m_{0} = |\ff m_{0}|$.
With this we have $\ff \eta \times \ff m_{0} = 0$ and thus
\be
\ff \eta = z_{K} \frac{\ff m_{0}}{m_{0}} 
\: .
\labeq{eta}
\ee
The total energy is minimized for the sign $z_{K} = - \mbox{sign} K = - K/|K|$ as argued below, see \refeq{heff}.
With Eq.\ (\ref{eq:coll}) the explicit constraint \refeq{con} finally reads as
\be
\ff n_{0,i} (\ff m) 
= 
z_{i} z_{K} \frac{\sum_{r} z_{i_{r}} \ff m_{r}}{|\sum_{r} z_{i_{r}} \ff m_{r} |} 
\: .
\label{eq:n0m}
\ee

Note that the function $\ff n_{0,i}(\ff m)$ is singular for $\ff m_{0}=\sum_{r} z_{i_{r}} \ff m_{r}  = 0$. 
The condition $m_{0}=0$ specifies a submanifold embedded in the full configuration space.
Though this has zero measure, we have to keep this in mind and must exclude trajectories $\ff m(t)$ crossing the submanifold. 
More importantly, one cannot choose initial conditions with $m_{0}(t=0)=0$ within the ASD theory. 
Consider the case of two impurity spins $R=2$ as an example.
Since $z_{i_{1}} = +1$ is already fixed (see above), there are two cases to be taken into account: $z_{i_{2}} = \pm 1$. 
For $z_{i_{2}} = - z_{i_{1}} = -1$, the parallel configuration $\ff m_{1} = \ff m_{2}$ is singular, while for $z_{i_{2}} = z_{i_{1}} = +1$ the antiparallel configuration $\ff m_{1} = - \ff m_{2}$ is singular must be excluded.
The physical meaning of the singularity becomes obvious at a later stage (see Sec.\ \ref{sec:res}). 

\section{Effective Hamiltonian and naive adiabatic theory}
\label{sec:eff}

Having the explicit form of the constraint, Eq.\ (\ref{eq:n0m}), at hand, we proceed with derivation of the effective Hamiltonian $H_{\rm eff}(\ff m)$ for arbitrary $R$.
This is obtained from $H(\ff n, \ff m)$ by using the constraint to eliminate the host-spin degrees of freedom, 
which yields $H_{\rm eff}(\ff m) = H(\ff n_{0}(\ff m),\ff m)$. 
Concretely, for the considered case of a collinear host-spin configuration, \refeq{coll}, we have
\be
H_{\rm eff}(\ff m)
=
E_{0}
+
K s S\sum_{r=1}^{R}  z_{i_{r}} \ff \eta \ff m_{r}
-
S \sum_{r=1}^{R} \ff m_{r} \ff B_{r} \: , 
\ee
where $E_{0} = (s^{2}/2) \sum_{i,i'=1}^{L} J_{ii'} z_{i} z_{i'}$ is the $\ff m$-independent ground-state energy of the host-spin Hamiltonian $H_{\rm host}(\ff n)$.
With $\ff \eta = z_{K} \ff m_{0} / m_{0}$, see \refeq{eta}, we find
\be
H_{\rm eff}(\ff m)
=
E_{0}
+
z_{K} K s S m_{0}
-
S \sum_{r=1}^{R} \ff m_{r} \ff B_{r}
\: .
\labeq{heff}
\ee
Note that the total energy is at a minimum for $z_{K} K = - |K| < 0$. 
This justifies the above choice $z_{K} = - \mbox{sign} K$.

As mentioned before, it is very tempting to derive the slow-spin dynamics solely from this effective Hamiltonian, see \refeq{na} and the related discussion.
This constitutes the naive adiabatic theory. 
The corresponding equations of motion of the naive theory are easily derived. 
We note that $\partial m_{0} / \partial \ff m_{r} = z_{i_{r}} \ff m_{0} / m_{0}$ and find
\be
\frac{\partial H_{\rm eff}(\ff m)}{\partial \ff m_{r}} 
=
- |K| s S 
z_{i_{r}} \frac{\ff m_{0}}{m_{0}}
- 
S \ff B_{r}
\: .
\ee
This yields
\be
\dot{\ff m}_{r} = \frac{1}{S} \frac{\partial H_{\rm eff}(\ff m)}{\partial \ff m_{r}} \times \ff m_{r} 
=
\left(
- |K|  s 
z_{i_{r}} \frac{\ff m_{0}}{m_{0}} 
-  
\ff B_{r} 
\right)
\times \ff m_{r}
\: 
\ee
or
\be
\dot{\ff m}_{r} 
=
- \frac{|K| s}{m_{0}} z_{i_{r}} 
\sum_{r'}
z_{i_{r'}} \ff m_{r'}
\times \ff m_{r}
-  
\ff B_{r} \times \ff m_{r}
\: 
\labeq{ndyn}
\: .
\ee
This is a comparatively simple nonlinear system of $R$ differential equations of motion for the $R$ impurity spins. 

The naive adiabatic theory is in fact conceptually incorrect, as the constraint is directly used to simplify the Hamiltonian, which is generally not justified. 
Before we proceed with the (conceptually correct) adiabatic spin dynamics (ASD), however, let us discuss some special cases and consequences of the naive theory.

For $\ff B_{r}=0$, the effective slow impurity-spin dynamics takes place on the time scale set by $1/K$, as is obvious from \refeq{ndyn}. 
One also easily verifies that the total energy $H_{\rm eff}$ and the total impurity spin $S \ff m_{\rm tot} \equiv S \sum_{r} \ff m_{r}$, and also $m_{0}$ are conserved quantities. 
Multiplying \refeq{ndyn} with $z_{i_{r}}$ and summing over $r$, we see that 
$\ff m_{0}$ precesses around $\ff m_{\rm tot}$:
\be
\dot{\ff m}_{0} = \frac{|K| s}{m_{0}} \ff m_{\rm tot} \times \ff m_{\rm 0}
\labeq{m0dyn}
\ee
with frequency
\be
\omega_{\rm p} = |K| s m_{\rm tot} / m_{0} \: . 
\labeq{pf0}
\ee

For two impurity spins, $R=2$, we have $\ff m_{\rm tot} = \ff m_{1} + \ff m_{2}$, and with $\ff m_{\rm tot}$ the angle $\vartheta$ that is enclosed by $\ff m_{1}$ and $\ff m_{2}$ is conserved as well. 
In this case, as is directly seen from \refeq{ndyn}, $\ff m_{1}$ and $\ff m_{2}$ precess with the same frequency $\omega_{\rm p}$ around $\ff m_{\rm tot}$.

The precession frequency decisively depends on the geometry.
Let us assume that $z_{i_{1}} = - z_{i_{2}}$ (note that we fixed $z_{i_{1}} = +1$). 
This happens to be the case, e.g., for an antiferromagnetic ground-state configuration of the host spins, $\ff n_{0,i} = \pm (-1)^{i} \ff \eta$, if the distance $i_{1}-i_{2}$ between the two impurity spins is odd. 
Then $\ff m_{0} = \ff m_1 - \ff m_{2}$ and thus 
$m_{0} = \sqrt 2 \sqrt{1 - \cos \vartheta}$. 
With $m_{\rm tot} = \sqrt 2 \sqrt{1+\cos \vartheta}$ we find
\be
  \omega_{\rm p}
  = 
  |K| s \sqrt{\frac{1+\cos \vartheta}{1-\cos \vartheta}}
  = 
  |K| s \cot (\vartheta/2) \: .
\labeq{om0}
\ee
For $z_{i_{1}} = + z_{i_{2}} = +1$, on the other hand, we have $\ff m_{0} = \ff m_{\rm tot}$ by definition, and \refeq{m0dyn} is useless.
The same holds for arbitrary $R$ and $z_{i_{r}} = + 1$ for {\em all} $r$, such that again $\ff m_{0} = \ff m_{\rm tot}$.
Going back to \refeq{ndyn}, we have $\dot{\ff m}_{r} = - (|K|s/m_{\rm tot}) \ff m_{\rm tot} \times \ff m_{r}$ in this case, i.e., each impurity spin precesses with frequency 
\be
\omega_{\rm p}
=
|K|s
\labeq{om0t}
\ee
around the total impurity spin.

Let us finally emphasize that the naive theory violates total spin conservation (for $\ff B_{r}=0$). 
The total impurity spin $\ff S_{\rm tot} = S \ff m_{\rm tot}$ is a constant of motion as stated above. 
The total host spin $\ff s_{\rm tot} = \sum_{i} \ff s_{i} = s \sum_{i} z_{i} \ff \eta$, however, has a nontrivial precession dynamics in the case $z_{i_{1}} = - z_{i_{2}}$ and for $\Delta \equiv \sum_{i} z_{i} \ne 0$ or, equivalently, for $\ff s_{\rm tot} = s \Delta \ff \eta \ne 0$.
Hence, $\ff S_{\rm tot} + \ff s_{\rm tot}  \ne \mbox{const}$.
This shortcoming is cured by the adiabatic spin dynamics theory, see below.

\section{Topological charge density and spin torque}
\label{sec:top}

In addition to the term on the right-hand side of \refeq{ndyn}, there is an additional contribution to the equations of motion of the ASD resulting from the topological spin torque, see \refeq{asd}. 
To derive this contribution, we must calculate the topological charge density \refeq{cd} from the explicit form of the constraint, Eq.\ (\ref{eq:n0m}). 

This is done straighforwardly. 
First, we have:
\be
\frac{\partial n_{0,i\mu}(\ff m) }{\partial m_{r'\mu'}}
=
z_{i} z_{K} z_{i_{r'}} \frac{1}{m_{0}} \left( 
\delta_{\mu\mu'} - \frac{m_{0\mu} m_{0\mu'}}{m_{0}^{2}}
\right) \: , 
\labeq{dn0dm}
\ee
where the $\ff m$ dependence is due to $\ff  m_{0} = \sum_{r} z_{i_{r}} \ff m_{r}$.
Inserting this expression in Eq.\ (\ref{eq:cd}), 
\ba
e^{(i)}_{r\mu,s\nu}(\ff m)
&=&
\frac{1}{4\pi}
\sum_{\rho\sigma\tau}
\varepsilon_{\rho\sigma\tau}
z_{i} z_{K} z_{i_{r}} \frac{1}{m_{0}} \left( 
\delta_{\rho\mu} - \frac{m_{0\rho} m_{0\mu}}{m_{0}^{2}}
\right)
\nonumber \\
&\times&
z_{i} z_{K} z_{i_{s}} \frac{1}{m_{0}} \left( 
\delta_{\sigma\nu} - \frac{m_{0\sigma} m_{0\nu}}{m_{0}^{2}}
\right)
z_{i} z_{K} \frac{m_{0\tau}}{m_{0}}
\: ,
\nonumber \\
\ea
expanding, exploiting the condition (\ref{eq:equi}), and carrying out the sums over $\rho, \sigma$, we find
\be
e^{(i)}_{r\mu,s\nu}(\ff m)
=
\frac{1}{4\pi}
z_{i} z_{K} z_{i_{r}} z_{i_{s}} 
\sum_{\tau}
\varepsilon_{\mu\nu\tau}
m_{0\tau}
\frac{1}{m^{3}_{0}} 
\: .
\ee
Summation over $i$ yields the tensor field defined in \refeq{omdef}:
\be
  \Omega_{r\mu,s\nu}(\ff m) 
  = 
  z_{i_{r}} z_{i_{s}} z_{K} s \Delta 
  \sum_{\tau} \varepsilon_{\mu\nu\tau} m_{0\tau} \frac{1}{m^{3}_{0}} \: .
\labeq{omres}
\ee  
Here, we have defined 
\be
\Delta \equiv \sum_{i} z_{i}
\: . 
\ee
We see that, in case of a collinear host-spin ground state, \refeq{coll}, a nonzero field $\Omega_{r\mu,s\nu}(\ff m)$ requires a ground state with a finite total host-spin magnetization $\ff n_{\rm tot} = \sum_{i} \ff n_{0,i} = \sum_{i} z_{i} \ff \eta = \Delta \, \ff \eta$. 
Its modulus $n_{\rm tot} = \Delta$ is nonzero, for instance, in case of ferromagnetic exchange couplings $J<0$, or in case of antiferromagnetic couplings $J>0$ when $L$ is odd.

Inserting the result for the tensor field into \refeq{tj} yields:
\ba
  \ff T_{r} &=& \sum_{\mu} T_{r\mu}(\ff m, \dot{\ff m}) \ff e_{\mu}
  = \sum_{s\mu\nu} \Omega_{r\mu,s\nu}(\ff m) \dot{m}_{s\nu} \ff e_{\mu}
  \nonumber \\
  &=& 
  z_{i_{r}} \frac{s \Delta}{m^{3}_{0}} \sum_{s} 
  z_{i_{s}} \sum_{\mu\nu\tau} 
  \varepsilon_{\mu\nu\tau} m_{0\tau}  \dot{m}_{s\nu} \ff e_{\mu}
  \nonumber \\
  &=& 
  z_{i_{r}} z_{K} \frac{s\Delta}{m^{3}_{0}} \dot{\ff m}_{0} \times \ff m_{0}  
  \: ,
\label{eq:tjcoll}  
\ea
so that the additional topological spin torque on the right-hand side of \refeq{asd} reads as:
\be
  \ff T_{r} \times \ff m_{r} =   z_{i_{r}} z_{K} \frac{s\Delta}{m^{3}_{0}} (\dot{\ff m}_{0} \times \ff m_{0}) \times \ff m_{r} 
\: .
\labeq{sumtm}
\ee
Note that the torque is independent of the coupling parameters and depends on the system geometry only.
Combining this with \refeq{asd} and \refeq{ndyn}, we arrive at the ASD equations of motion:
\be
\dot{\ff m}_{r} 
\!=\!
\left( \!
-  z_{i_{r}} \frac{|K|  s}{m_{0}} 
\ff m_{0}
+
z_{i_{r}} z_{K} \frac{s\Delta}{S m^{3}_{0}} \dot{\ff m}_{0} \! \times \! \ff m_{0}
-  
\ff B_{r} \!
\right)
\! \times  \ff m_{r}
\, .
\labeq{tdyn}
\ee
The first term is the same as in the naive approach, the second one is due to the topological spin torque.

\section{Adiabatic spin dynamics}
\label{sec:dis}

For the discussion of the equations of motion of adiabatic spin dynamics, see \refeq{tdyn}, we will consider several cases. 
Let us first check the case $R=1$, $K>0$, $J>0$, $i_{1} = 1$, and odd $L$, i.e., there is a single impurity spin only, $\ff m \equiv \ff m_{1}$, which couples antiferromagnetically at the first site of an antiferromagnetic host with an odd number of sites. 
We have $\ff m_{0} = \ff m_{\rm tot} = \ff m$ and $\Delta = +1$.
The naive equation of motion, \refeq{ndyn}, reads $\dot{\ff m} = \ff m \times \ff B$, and thus predicts precession around the external magnetic field with Larmor frequendy $\omega_{\rm p} = B$. 
Including the additional topological spin torque, however, we find
$S \dot{\ff m} = S \ff m \times \ff B - s (\dot{\ff m} \times \ff m) \times \ff m 
= S \ff m \times \ff B + s \dot{\ff m}$. 
This can be written in the form of the Landau-Lifschitz equation but leads to an anomalous precession frequency
\be
\omega_{\rm p} = \frac{B}{1-s/S} \: .
\labeq{anoom}
\ee
This is precisely the result derived in Ref.\ \cite{EMP20}.

Next we discuss constants of motion for the general case (but we assume $\ff B_{r}=0$).
We start by checking that the equation of motion respects the conservation of $|\ff m_{r}|=1$. 
This is the case (also for finite $\ff B_{r}$) since $\ff m_{r} \dot{\ff m}_{r} = 0$, see Eqs.\ (\ref{eq:asd}) or (\ref{eq:tdyn}).

Total energy energy conservation is ensured on general grounds as the equation of motion is derived within the standard Lagrange formalism and employing a scleronomic holonomic constraint. 
This has also been proven explicitly and discussed in detail in Ref.\ \cite{EMP20}.

Conservation of the total spin, i.e., the sum of the total impurity spin $\ff S_{\rm tot} = S \sum_{r} \ff m_{r}$ and the total host spin $\ff s_{\rm tot} = s \sum_{i} \ff n_{0,i}(\ff m)$, can be shown for the case of an SO(3) symmetric effective Hamiltonian $H_{\rm eff}(\ff m)$. 
This is detailed in Appendix \ref{sec:spincons}.

Conservation of the total impurity spin $\ff m_{\rm tot} = \sum_{r} z_{i_{r}} \ff m_{r}$ is not expected in general. 
Summing both sides of \refeq{tdyn} over $r=1,...,R$ we get (for $\ff B_{r}=0$):
\be
\dot{\ff m}_{\rm tot}
=
z_{K} \frac{s\Delta}{S m^{3}_{0}} (\dot{\ff m}_{0} \times \ff m_{0} ) \times \ff m_{0}
\: , 
\labeq{mtdyn0}
\ee
which is nonzero for a finite topological spin torque, i.e., for $\Delta \ne 0$.
Note that this immediately implies $\dot{\ff m}_{\rm tot} \ff m_{0} = 0$.

Summing both sides of \refeq{tdyn} over $r$ after multiplying with $z_{i_{r}}$, we can derive an equation for the ``staggered'' sum of the impurity spins $\ff  m_{0} = \sum_{r} z_{i_{r}} \ff m_{r}$. 
For $\ff B_{r}=0$ we get:
\be
\dot{\ff m}_{0}
=
- 
\frac{|K|  s}{m_{0}} \ff m_{0} \times \ff m_{\rm tot}
+
z_{K} \frac{s\Delta}{S m^{3}_{0}} (\dot{\ff m}_{0} \times \ff m_{0}) \times \ff m_{\rm tot}
\: .
\labeq{msdyn}
\ee
We immediately have $\dot{\ff m}_{0} \ff m_{\rm tot} = 0$. 
Together with the above relation $\dot{\ff m}_{\rm tot} \ff m_{0} = 0$, this implies that the inner product $\ff m_{0} \ff m_{\rm tot}$ is conserved. 
Furthermore, we have
$(\dot{\ff m}_{0} \times \ff m_{0}) \times \ff m_{\rm tot}
=
- \dot{\ff m}_{0} (\ff m_{\rm tot} \ff m_{0})$ and therewith we can convert \refeq{msdyn} into an explicit differential equation:
\be
\dot{\ff m}_{0}
=
\frac{1}{
1+z_{K} \frac{s\Delta}{S m^{3}_{0}} \ff m_{0} \ff m_{\rm tot}
}
\frac{|K|  s}{m_{0}} \ff m_{\rm tot} \times \ff m_{0} 
\: .
\labeq{msexpl}
\ee
Multiplying both sides of the equation with $\ff m_{0}$, yields $\ff m_{0} \dot{\ff m}_{0} = 0$ and hence $m_{0} = \mbox{const}$ (if $\ff B_{r}=0$). 
Hence, the prefactor of $\ff m_{0} \times \ff m_{\rm tot}$ in \refeq{msexpl} is a constant of motion.

With $\ff m_{0} \dot{\ff m}_{0} = 0$ we can also simplify the double cross product in \refeq{mtdyn0}:
\be
\dot{\ff m}_{\rm tot}
=
- z_{K} \frac{s\Delta}{S m_{0}} \dot{\ff m}_{0} 
\: .
\ee
Multiplying with $\ff m_{\rm tot}$ and using $\dot{\ff m}_{0} \ff m_{\rm tot} = 0$, we see that the norm of $\ff m_{\rm tot}$ is conserved. 
Furthermore, $\dot{\ff m}_{0}$ on the right-hand side can be eliminated using \refeq{msexpl}, such that we are finally left with:
\be
\dot{\ff m}_{\rm tot}
=
 z_{K} \frac{s\Delta}{S m_{0}}  
\frac{1}{
1+z_{K} \frac{s\Delta}{S m^{3}_{0}} \ff m_{0} \ff m_{\rm tot}
}
\frac{|K|  s}{m_{0}} \ff m_{0} \times \ff m_{\rm tot}
\: .
\labeq{mtdyn}
\ee

Summing up, for $\ff B_{r}=0$, we have, besides energy and total spin conservation, the following conserved quantities: 
\be
m_{0} = \mbox{const.} \; , \quad
m_{\rm tot} = \mbox{const.} \; , \quad
\ff m_{0} \ff m_{\rm tot} = \mbox{const.}
\labeq{cons}
\ee
Furthermore, there are two coupled nonlinear ordinary differential equations, \refeq{msexpl} and \refeq{mtdyn}, for $\ff m_{0}$ and $\ff m_{\rm tot}$.

There is a link to the naive adiabatic theory, namely in the topologically trivial case where $\Delta = 0$, i.e., where the topological spin torque vanishes. 
Here, the total impurity spin is conserved, $\dot{\ff m}_{\rm tot} = 0$, and according to \refeq{msexpl}, $\ff m_{0}$ precesses around $\ff m_{\rm tot}$ with frequency $\omega_{\rm p} = |K| s m_{\rm tot} / m_{0}$. 
This precisely recovers the results of the naive theory, see \refeq{pf0}.

In the nontrivial case for $\Delta \ne 0$, an analytical solution of \refeq{msexpl} and \refeq{mtdyn} is obtained easily. 
To this end, we rewrite the equations as
\be 
\dot{\ff m}_{0}
=
c_{0} \ff m_{\rm tot} \times \ff m_{0} \; , \quad
\dot{\ff m}_{\rm tot}
=
c_{\rm tot} \ff m_{0} \times \ff m_{\rm tot} \; ,
\ee
with constants $c_{0}$ and $c_{\rm tot}$, and employ a scaling transformation to new variables $\widetilde{\ff m}_0 = \alpha_{0} \ff m_{0}$ and $\widetilde{\ff m}_{\rm tot} = \alpha_{\rm tot} \ff m_{\rm tot}$ such that the prefactors in the transformed equations of motion are equal (this is the case for $z_{K} \Delta > 0$) or differ by a sign only ($z_{K} \Delta <0$). 
Details are given in Appendix \ref{sec:ode}.
We find that both, $\ff m_{0}$ and $\ff m_{\rm tot}$, are precessing around the conserved total spin 
$\ff s_{\rm tot} + \ff S_{\rm tot}$, and the corresponding precession frequency is:
\be
  \omega_{\rm p}
  =
  \sqrt{
  c_{\rm tot}^{2} m_{0}^{2} 
  + 
  c_{0}^{2} m_{\rm tot}^{2} 
  \pm 2 \left| c_{0} c_{\rm tot} \right| \, \ff m_{0} \ff m_{\rm tot}
  }
  \: .
\labeq{om}
\ee
With the solutions $\ff m_{0}$, $\ff m_{\rm tot}$ at hand, the dynamics of the individual impurity moments $\ff m_{r}$ can be obtained from a numerical solution of \refeq{tdyn} for each $\ff m_{r}$ {\em separately}.
This situation is different from but reminiscent of gyroscope theory, since $\ff m_{r}$ precesses around a momentary axis specified by $\ff m_{0}$ and $\dot{\ff m}_{0}$, while $\ff m_{0}$ itself is precessing around an axis fixed in space. 
 
Finally, we consider the special case of two impurity spins $R=2$ and vanishing external fields $\ff B_{r}=0$. 
Here, it is sufficient to analyze the two coupled equations \refeq{msexpl} and \refeq{mtdyn}, rather than reverting to \refeq{tdyn} again, since the dynamics of $\ff m_{1}$ and $\ff m_{2}$ is fully determined via $\ff m_{\rm tot} = \ff m_{1} + \ff m_{2}$ and $\ff m_{0} = z_{i_{1}} \ff m_{1} + z_{i_{2}} \ff m_{2}$.

For the case $z_{i_{1}} = z_{i_{2}} = 1$, we have $\ff m_{0} = \ff m_{\rm tot}$, such that \refeq{msexpl} reduces to \refeq{mtdyn}.
\refeq{mtdyn} implies that the total impurity spin is conserved, $\dot{\ff m}_{\rm tot} = 0$, and thus the topological spin torque vanishes.
This is the topologically trivial case. 
Both impurity spins precess around $\ff m_{\rm tot}$ with frequency $\omega_{\rm p} = |K| s$, as discussed above, see \refeq{om0t}.

For the nontrivial case $z_{i_{1}} = - z_{i_{2}} = 1$, we have $\ff m_{0} = \ff m_{1} - \ff m_{2}$.
This implies $\ff m_{0} \ff m_{\rm tot} = (\ff m_{1} - \ff m_{2})(\ff m_{1}+\ff m_{2})=0$, since $\ff m_{1}$ and $\ff m_{2}$ are unit vectors, such that the last statement of \refeq{cons} becomes trivial. 
Further, the first and second one imply $\ff m_{1} \ff m_{2} = \cos \vartheta = \mbox{const}$.
Therewith, the equations of motion for $\ff m_{0}$ and $\ff m_{\rm tot}$ simplify and read: 
\be
\dot{\ff m}_{0}
=
\frac{|K|  s}{m_{0}} \ff m_{\rm tot} \times \ff m_{0} 
\: 
\labeq{m02}
\ee
and
\be
\dot{\ff m}_{\rm tot}
=
 z_{K} \frac{s\Delta}{S m_{0}}  
\frac{|K|  s}{m_{0}} \ff m_{0} \times \ff m_{\rm tot}
\: .
\labeq{mt2}
\ee
The precession frequency is:
\be
  \omega_{\rm p} = |K| s \, \frac{1}{m_{0}} 
  \sqrt{
  \frac{s^2 |\Delta|^2}{S^2} 
  +
  m_{\rm tot}^{2}
  } \: .
\ee
Assuming that $s=S$ and that $|\Delta|=1$ (antiferromagnetic host-spin configuration and odd $L$), we have (see Appendix \ref{sec:ode}):
\be
  \omega_{\rm p} 
  =
  |K| s\, \frac{\sqrt{1 + m_{\rm tot}^{2}}}{m_{0}}
   =
  |K| s \, \frac{1}{2\sin\nicefrac{\vartheta}2}
 \sqrt{1 + 4  \cos^2 \frac{\vartheta}2}
  \: .
  \labeq{pf}  
\ee
This also applies to the individual impurity spins: 
For $R=2$, the dynamics of the impurity spins, $\ff m_{1}$ and $\ff m_{2}$, is simple: 
They precess around the axis specified by the total spin with the same frequency as $\ff m_{0}$ and $\ff m_{\rm tot}$.

Note that the precession frequency approaches $\omega_{\rm p} \to |K| s / 2$ for $\vartheta \to \pi$, i.e., when one approaches the global ground state.
This must be compared with the result $\omega_{\rm p} \to 0$ that is obtained by the naive adiabatic theory. 
For $\vartheta \to 0$, on the other hand, i.e., for $m_{0} \to 0$, the precession frequency diverges as $\omega_{\rm p} \to \sqrt{5} |K| s / \vartheta$. 
This divergence originates from the above-mentioned singularity, cf.\ the discussion following Eq.\ (\ref{eq:n0m}).

We would like to emphasize that already the simple $R=2$ case demonstrates that an effective impurity-spin dynamics is not Hamiltonian. 
There is no effective Hamilton function $H_{\rm eff}(\ff m_{1},\ff m_{2})$ with which the equations of motion (\ref{eq:m02}) and (\ref{eq:mt2}) can be reproduced. 
Any nontrivial two-impurity-spin Hamiltonian model would be of the form $H_{\rm eff}(\ff m_{1},\ff m_{2})
 = K f(\ff m_{1} \ff m_{2})$ with some smooth real function $f$, which would immediately, and incorrectly, imply that $\ff m_{1}+\ff m_{2}$ is conserved.

\section{Numerical results}
\label{sec:res}

\begin{figure}[t]
\includegraphics[width=0.55\columnwidth]{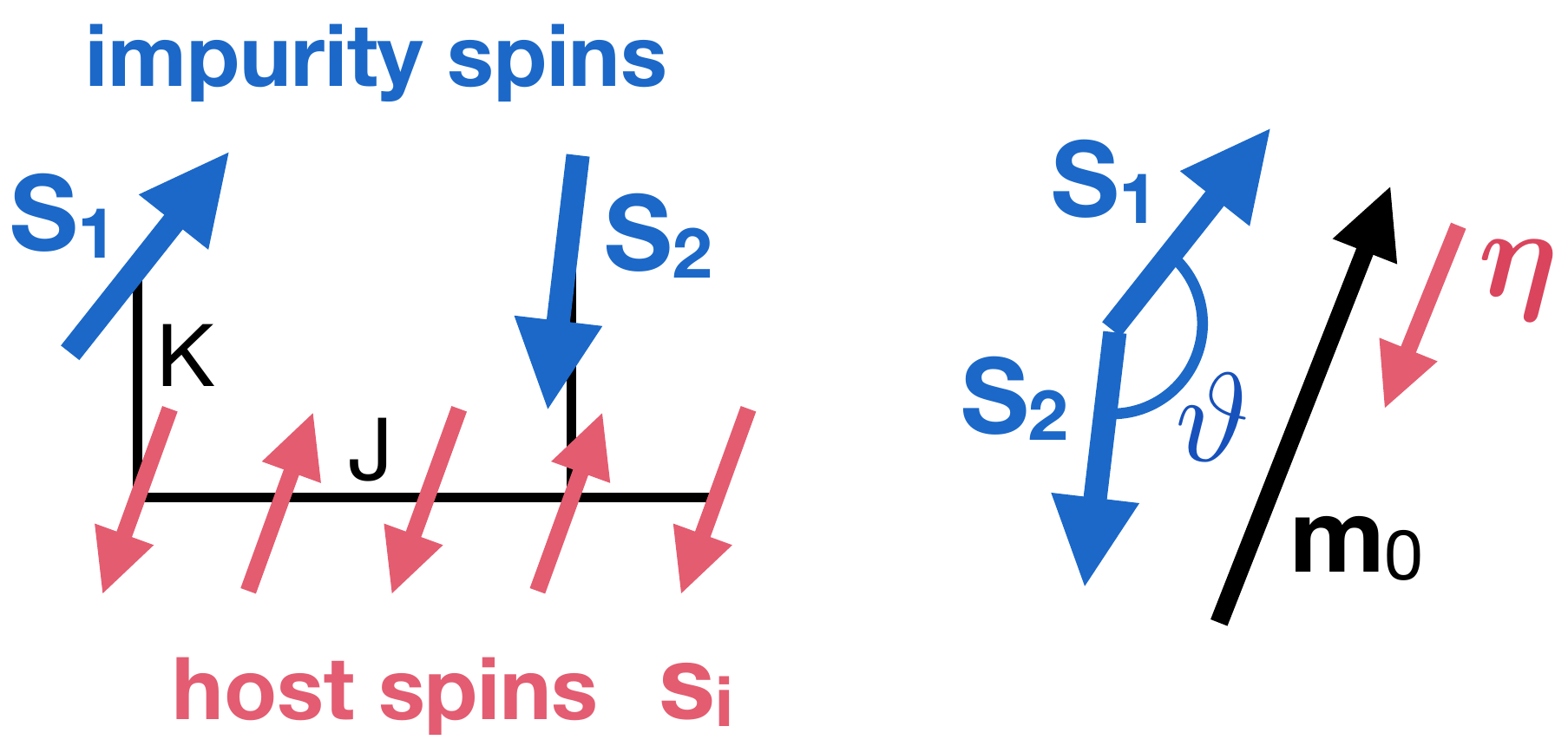}
\caption{
System geometry: $L=5$ host spins (length $s=1$), $R=2$ impurity spins ($S=1$) coupled to the host spins at sites $i_{1}=1$, $i_{2}=4$, antiferromagnetic exchange $J,K>0$, hence: $z_{i_{1}}=+1$, $z_{i_{2}}=-1$, and $\Delta=1$. 
Initial angle enclosed by the two impurity spins: $\vartheta$.
Initially the host-spin system is in its antiferromagnetic ground state for the given impurity-spin configuration, i.e., $\ff \eta = - \ff m_{0}/m_{0}$.
}
\label{fig:1d}
\end{figure}

It remains to check the validity of the adiabatic spin-dynamics theory, i.e., to find out in which parameter regime the adiabatic approximation is justified. 
We therefore compare the predictions of the ASD with the numerical solution of the full set of equations of motion (\ref{eq:hameq}). 
For the sake of simplicity, we first pick a geometry with $L=5$ host spins and $R=2$ impurity spins with antiferromagnetic couplings $J,K>0$, see Fig.\ \ref{fig:1d}.
The impurity spins are coupled to the host at sites $i_{1}=1$ and $i_{2}=4$. 
Since $L$ is odd and $z_{i_{1}} = - z_{i_{2}} = 1$, this is a realization of the topologically nontrivial case.

\begin{figure*}[t]
\includegraphics[width=1.95\columnwidth]{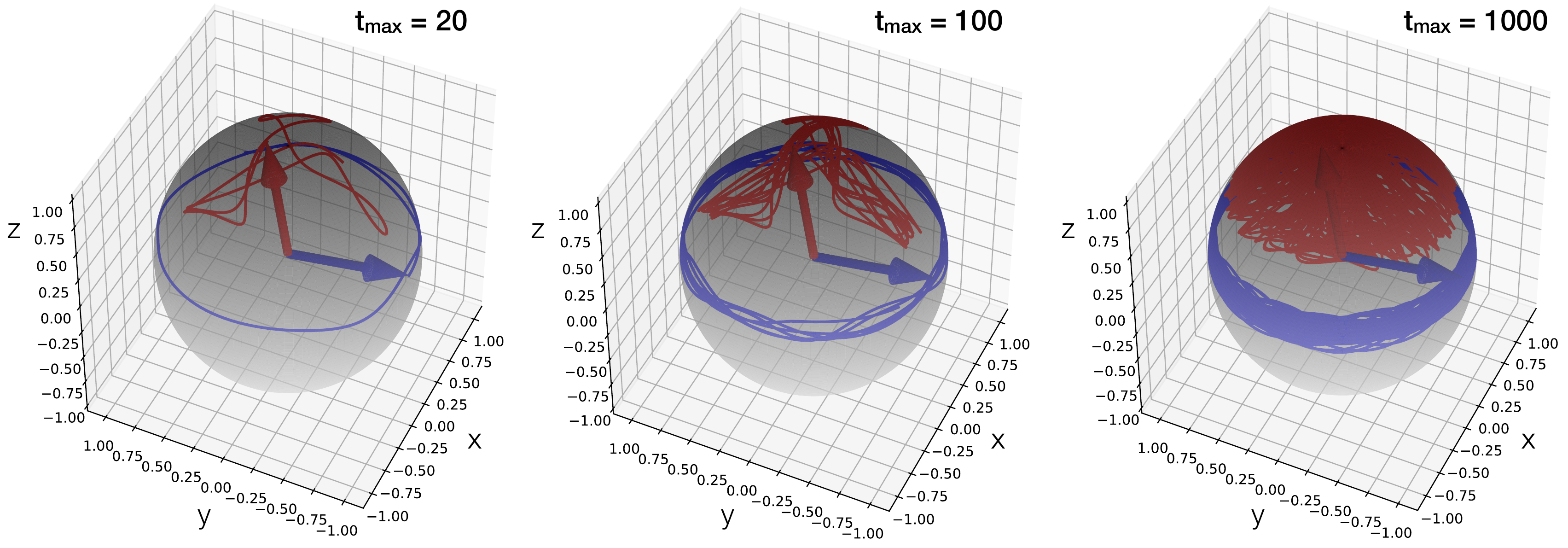}
\caption{
Trajectories of the two impurity spins on the Bloch sphere as obtained by numerical solution of the full set of equations of motion (\ref{eq:hameq}) for $K=J$, $\vartheta=\pi/2$ for different maximal propagation times $t_{\rm max}$ as indicated (in units of $K^{-1}$).
At time $t=0$ the system is prepared as indicated by the arrows.
$\ff m_{1}$: blue, $\ff m_{2}$: red.
The host spins are in their ground-state configuration for given impurity spins. 
The total spin is parallel to the $z$-axis.
}
\label{fig:ch}
\end{figure*}


\begin{figure}[b]
\includegraphics[width=0.65\columnwidth]{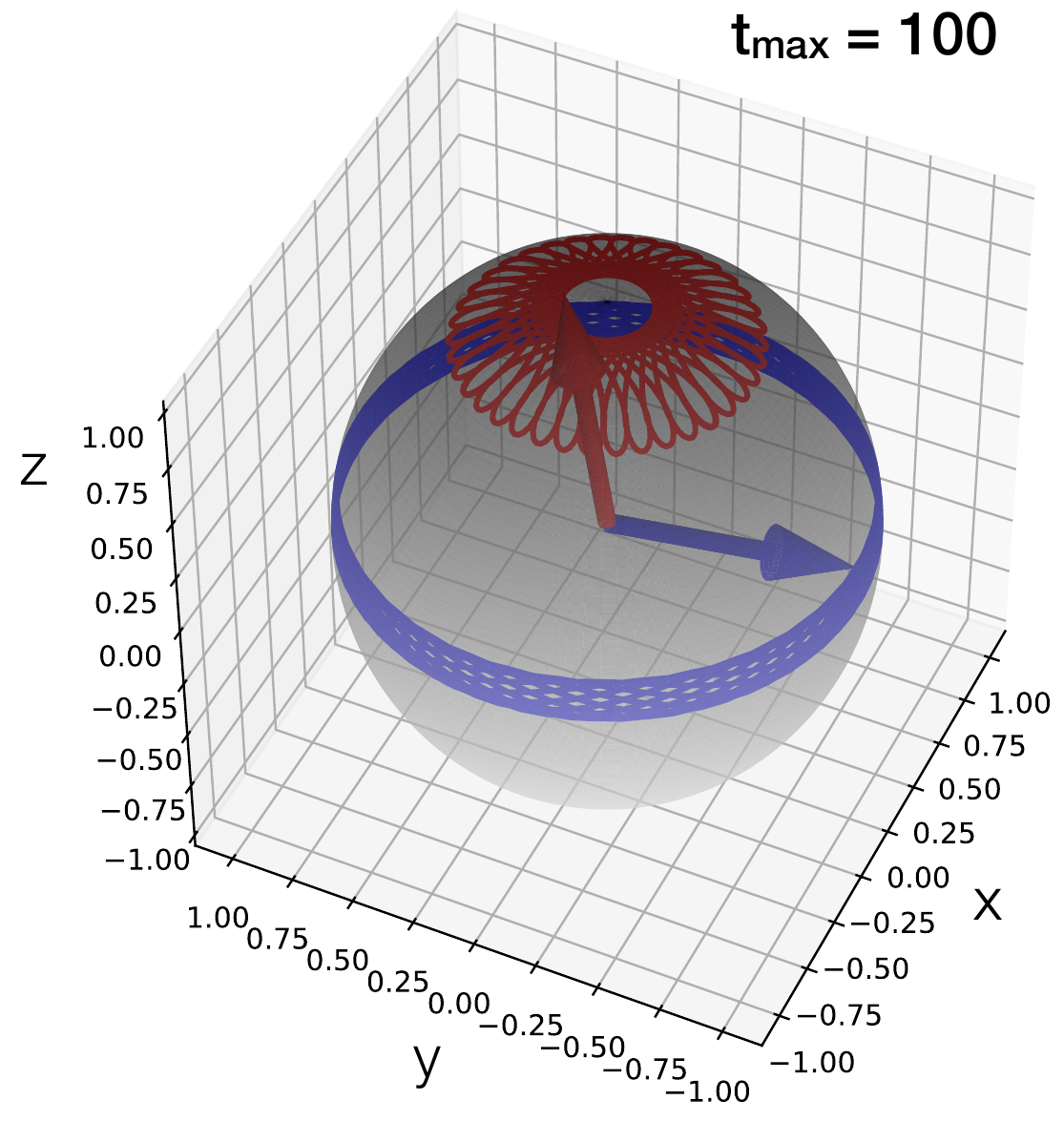}
\caption{
The same as Fig.\ \ref{fig:ch} but for $K/J = 10^{-5}$. Maximal propagation time $t_{\rm max}=100 K^{-1}$.
}
\label{fig:chad}
\end{figure}

With Fig.\ \ref{fig:ch} we give an example result which is characteristic of the real-time spin dynamics if $K$ and $J$ are of the same order of magnitude, and if the initial configurations of impurity spins is far from the (antiferromagnetic) ground state configuration. 
The initial host-spin configuration is taken to be the ground-state configuration for the given impurity-spin directions. 
As is demonstrated with Fig.\ \ref{fig:ch}, we find an extremely complex dynamics as it is characteristic for a nonlinear classical Hamiltonian system with several degrees of freedom.
For long times, the trajectories cover the entire phase space that is accessible under total energy and total spin conservation. 

Clearly, adiabatic spin dynamics is only expected to be realized in the weak-coupling limit $K\ll J$. 
Fig.\ \ref{fig:chad} provides an example for the same setup and parameters as in Fig.\ \ref{fig:ch} but for $K/J=10^{-5}$.
The motion is mainly precessional but there is an additional nutation visible. 
This nutation effect is not captured by the ASD but gets weaker and finally almost disappears when the initial impurity-spin configuration is chosen closer and closer to a ground-state configuration. 

\begin{figure}[b]
\includegraphics[width=0.85\columnwidth]{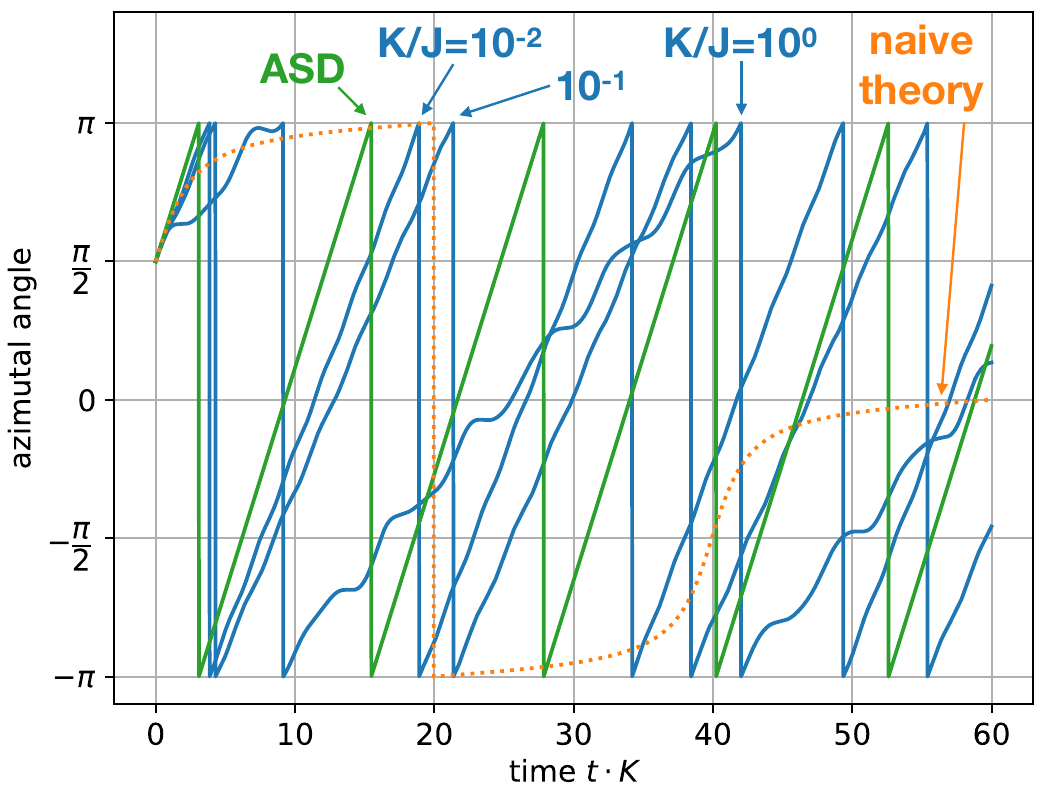}
\caption{
Time dependence (in units of $K^{-1}$) of the azimuthal angle (mod $2\pi$) in the precession dynamics of $\ff m_{0}$ around the conserved total spin for the geometry displayed in Fig.\ \ref{fig:1d} and for $\vartheta = 0.95 \pi$.
Numerical solution of the full set of equations of motion (\ref{eq:hameq}) for various ratios $K/J$ as indicated: blue.
Naive theory: orange.
Adiabatic spin dynamics (ASD): green.
}
\label{fig:azi}
\end{figure}

If the initially enclosed angle $\vartheta \approx \pi$, the dynamics is even more regular. 
Fig.\ \ref{fig:azi} displays an example, where $\vartheta = 0.95 \pi$ and where the host is in the corresponding ground state initially. 
Still, for $K/J=1$, the individual impurity spins $\ff m_{r}$ for $r=1,2$ show a rather complicated time evolution (not shown). 
The staggered sum $\ff m_{0} = \sum_{r} z_{i_{r}} \ff m_{r} = \ff m_{1} - \ff m_{2}$, on the other hand, is already much closer to a purely precessional motion.
The figure shows the time dependence of the azimuthal angle $\varphi$, modulo $2\pi$, of $\ff m_{0}$ with respect to the total conserved spin $\ff s_{\rm tot} + \ff S_{\rm tot}$. 
This azimuthal angle more or less grows linearly in time but with some weak additional structure superimposed.
With decreasing ratio $K/J$, the additional superimposed oscillations get weaker and weaker and are only hardly visible when $K/J=0.01$.

The green line in Fig.\ \ref{fig:azi} shows the result of the ASD theory, which predicts a purely precessional motion and correspondingly a linear increase of $\varphi$ as a function of $t$.
We see that with decreasing ratio $K/J$, the trajectory of $\ff m_{0}$, obtained from the full theory, appears to converge to the ASD result.
Not only the additional structure diminishes further and further but also the ASD prediction of the angular velocity $d\varphi / d t$ seems to be approached in the $K/J\to 0$ limit.

On the contrary, the prediction of the naive adiabatic theory, see \refeq{m0dyn}, which is directly obtained from the effective Hamiltonian, \refeq{heff}, is completely off. 
The approach does yield a purely precessional motion but mistakenly around the total impurity spin $\ff m_{\rm tot}$, which is a constant of motion within the naive theory but not within the ASD and the full theory. 
Furthermore, the angular velocity is by far too small or, as can be seen in the figure, the period $2\pi / \omega_{\rm p}$ is by far too large.

With decreasing $K/J$ also other quantities appear to converge to the predictions of the ASD (not shown).
We find, for example, that the modulus of the staggered and of the total impurity spin, $m_{0}$ and $m_{\rm tot}$, and the scalar product $\ff m_{0}\ff m_{\rm tot}$ or, equivalently, the enclosed angle $\vartheta$ approach constants when $K/J\to 0$, as is stated in \refeq{cons}.
Furthermore, also the dynamics of individual impurity spins seem to more and more approach a purely precessional motion with the same precession frequency.

\begin{figure}[t]
\includegraphics[width=0.85\columnwidth]{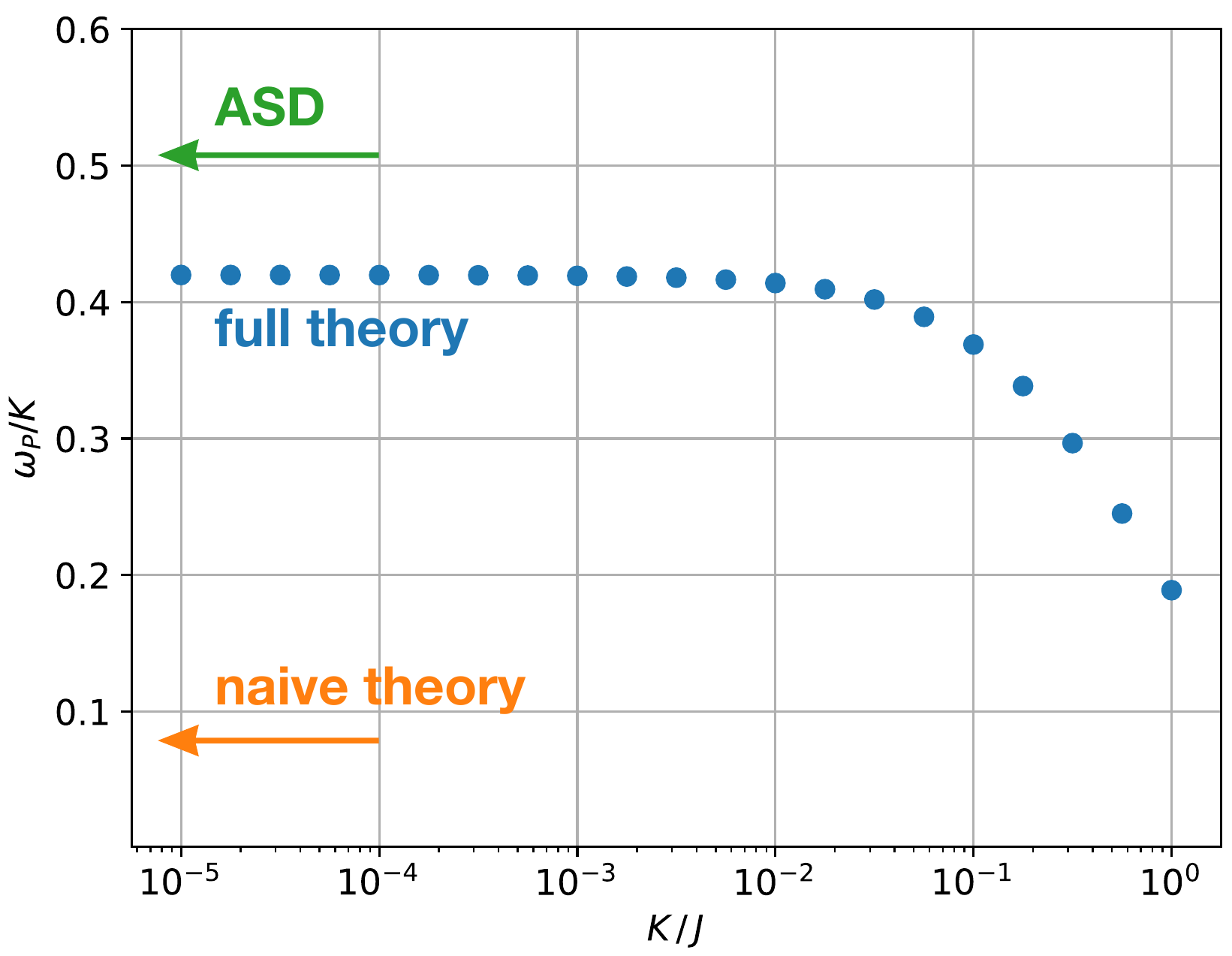}
\caption{
Precession frequency $\omega_{\rm p}$ of the staggered sum $\ff m_{0}$ as a function of the coupling strength $K/J$ at $\vartheta = 0.95\pi$. 
Results obtained from the numerical solution of the full set of equations of motion, \refeq{hameq}, compared to the predictions of the ASD and of the naive adiabatic spin dynamics in the limit $K/J\to 0$.
}
\label{fig:koverj}
\end{figure}

There is, however, a finite residual difference between the full theory, \refeq{hameq}, and the ASD persisting in the limit $K/J\to 0$.
This is demonstrated in Fig.\ \ref{fig:koverj} where the precession frequency in the real-time dynamics of $\ff m_{0}$ is plotted as a function of $K/J$. 
Since $\vartheta = 0.95\pi$, the impurity-spin configuration is close to a ground-state configuration, such that there is a frequency with dominant weight in the Fourier analysis of the data. 
This frequency smoothly depends on $K/J$ and approaches the frequency of the almost pure precessional dynamics that remains in the limit $K/J\to 0$. 
Already at $K/J \approx 10^{-3}$ it approaches saturation, although at a level that differs from the ASD result (green arrow) by about 15\%.
This implies that even in the weak-coupling limit, the host spins do not completely adiabatically follow the impurity-spin dynamics. 
The observation of a close-to-adiabaticity dynamics has already been made earlier for the single-spin ($R=1$) case \cite{SP15}. 
Turning to the naive adiabatic theory, see orange arrow in Fig.\ \ref{fig:koverj}, the predicted precession frequency is again by far too small, aside from the fact that the precession axis is predicted incorrectly.

\begin{figure}[t]
\includegraphics[width=0.95\columnwidth]{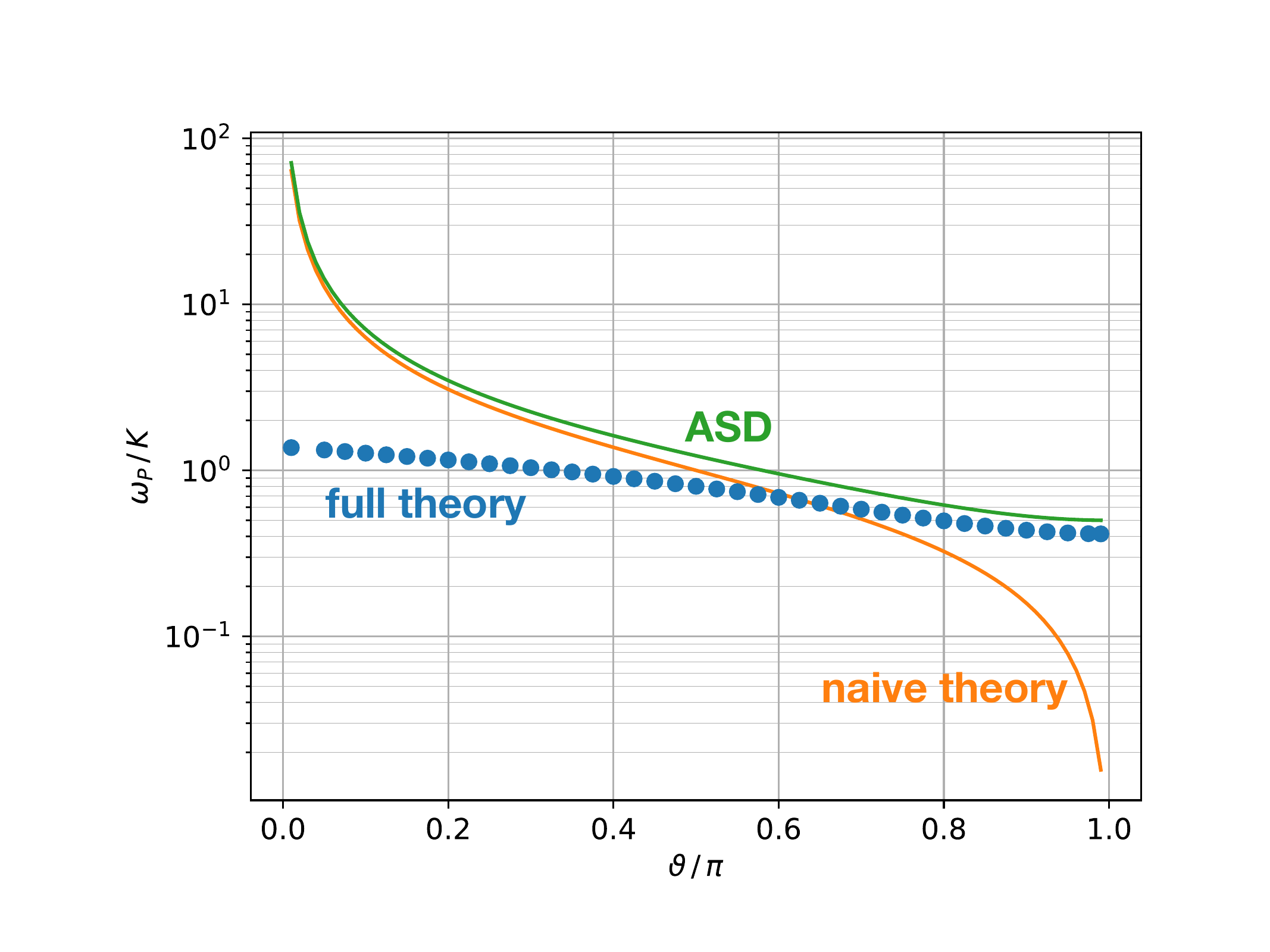}
\caption{
$\vartheta$ dependence of the precession frequency $\omega_{\rm p}$ as predicted by the ASD (green), see \refeq{pf}, and compared to the numerical data obtained at $K/J=10^{-5}$ by solving the full set of equations of motion (blue points), \refeq{hameq}, and to the naive theory (orange), \refeq{om0}.
}
\label{fig:theta}
\end{figure}

While the ASD theory is at least qualitatively correct at small $K/J$ and $\vartheta \to \pi$, it must break down for initial impurity-spin configurations that are far from a ground-state configuration. 
This is demonstrated with Fig.\ \ref{fig:theta}, where the analytical result (\ref{eq:pf}) for the ASD precession frequency is plotted against $\vartheta$ and compared to the numerical data for a coupling strength $K/J=10^{-5}$ deep in the weak-coupling limit. 
For $\vartheta \to \pi$ the ASD is close to the numerical data and correctly predicts a finite nonzero frequency $\omega_{\rm p}=|K|s/2$ in the limit, while there is a remaining discrepancy visible, as discussed above. 

With decreasing $\vartheta$ and increasing parametric distance to the ground state, however, the ASD is less reliable. 
This is understood easily and eventually results from the singular constraint, see
Eq.\ (\ref{eq:n0m}), on the $m_{0}=0$ manifold. 
For $\vartheta \to 0$, i.e., for $\ff m_{1} = \ff m_{2}$ initially, we have $\ff m_{0}=0$ initially, and thus $m_{0}=0$ at all times within the ASD. 
This singularity leads to the divergence of the frequency $\omega_{\rm p} \to \sqrt{5} |K| s / \vartheta$ for $\vartheta \to 0$, see \refeq{pf}.
It results from the fact that the ground-state host-spin configuration cannot be determined unambiguously for the maximally excited impurity-spin configuration, and that there is two-dimensional manifold of degenerate host-spin configurations in this case.

Vice versa, a divergent precession frequency implies a {\em fast} impurity-spin dynamics, i.e., a violation of the central assumption of a slow, adiabatic or close-to-adiabatic motion.
We note that this kind of inherently built-in breakdown of the theory is already known from the single-spin ($R=1$) case with a finite external magnetic field \cite{EMP20}, where it shows up, however, at a different point in parameter space, namely for $s \to S$, see \refeq{anoom}.

Finally, the naive adiabatic spin-dynamics theory is neither correct in the $\vartheta \to \pi$ nor in the $\vartheta \to 0$ limit. 
In the latter case, the precession frequency diverges as $\omega_{\rm p} \to 2|K|s / \vartheta$.

\begin{figure}[t]
\includegraphics[width=0.95\columnwidth]{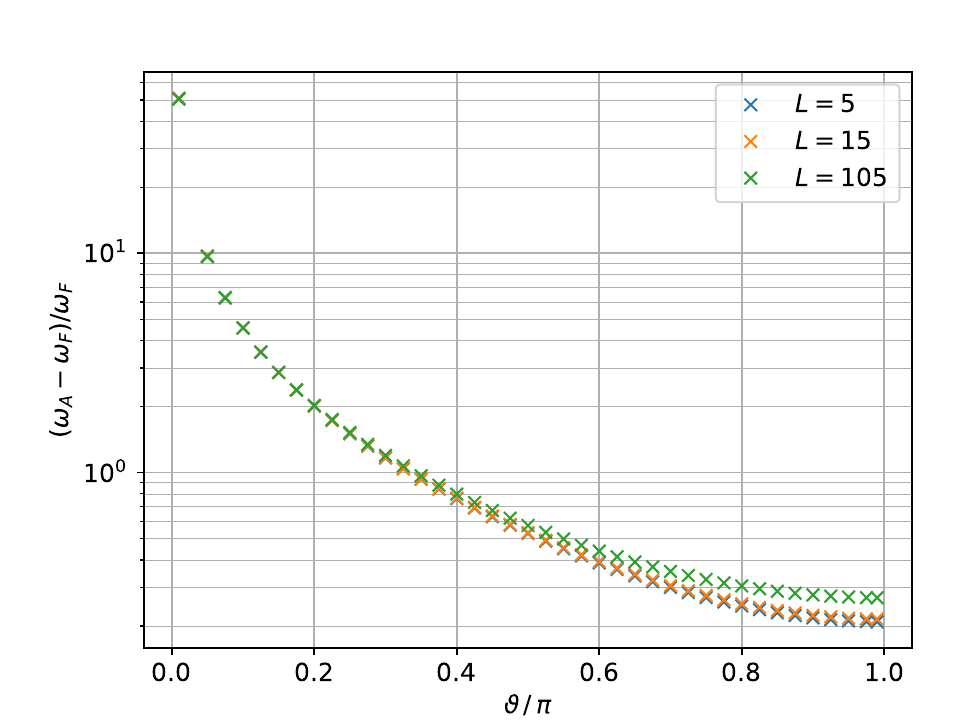}
\caption{
Normalized difference of the ASD precession frequency $\omega_{A}$ and the frequency $\omega_{F}$ obtained numerically from the full solution of \refeq{hameq} as function of $\vartheta$ at $K/J=10^{-3}$.
Results for various system sizes $L=5,15,105$. 
The impurity spins couple to positions $i_{1}=1$ and $i_{2}=L-1$.
}
\label{fig:tdlimit}
\end{figure}

The discussion of the results and the conclusions also apply to larger system sizes. 
This is demonstrated with Fig.\ \ref{fig:tdlimit}, where the normalized difference between the ASD precession frequency and the precession frequency of the numerical solution of the full set of Hamilton equations of motion is plotted against $\vartheta$ for different $L$. 
Since $L$ is odd in all cases and since the impurity spins are coupled to the host at $i_{1}=1$ and $i_{2}=L-1$, we have the topologically non-trivial case at hand.
We see that the difference diverges for $\vartheta \to 0$, as discussed above.
For $\vartheta \to \pi$, on the other hand, the residual difference becomes larger with increasing $L$. 
This means that it becomes more and more difficult to enforce close-to-adiabatic dynamics. 
Obviously, this is due to the necessity to communicate the relative impurity-spin configuration over large distances. 

\section{Beyond the adiabatic approximation}
\label{sec:bey}

One way to improve the theory and to go beyond the adiabatic approximation is to relax the constraint  \refeq{n0m} defining the ASD.
For the weak-coupling limit $K\ll J$, it is tempting to keep the host spins tightly coupled together but to relax the demand that the host-spin configuration should be given, at any instant of time, by the ground-state configuration for the currently present configuration of the impurity spins.
This idea can be formalized by substituting \refeq{n0m} by the constraint
\be
    \ff n_{i} 
    \stackrel{!}{=} 
    \ff n_{0,i}(\ff \eta)
    =
    z_{i} \ff \eta \: ,
\label{eq:tcon}
\ee
where $\ff \eta$ is a (three-component) dynamical degree of freedom normalized to unity, $\ff \eta^{2}=1$.
If we consider host spins on a bipartite lattice or, for the sake of simplicity, on a one-dimensional chain of sites $i=1,...,L$ that are tightly bound together via a strong antiferromagnetic coupling $J>0$, we have $z_{i}=(-1)^{i+1}$, with the convention $z_{1}=+1$.

A conceptual disadvantage of an effective spin-dynamics theory under these tight-binding constraints is that it necessarily involves (with $\ff \eta$) dynamical host degrees of freedom, such that one will not end up with an effective theory of the impurity-spin degrees of freedom only. 
Clearly, this is the price to be paid when aiming at an improved theory beyond the ASD.
On the other hand, a formal advantage is that there is no singularity and that no submanifold of spin configurations must be excluded, as compared to the ASD, cf.\ the discussion following \refeq{n0m}.

Again, one must be very careful when imposing the constraint \refeq{tcon}.
In Appendix \ref{sec:tb}, it is demonstrated that one runs into unacceptable inconsistencies, if one attempts to use the constraint (\ref{eq:tcon}) directly for a simplification of the full set of equations of motion (\ref{eq:hameq}).
The proper way is rather to start from the action principle again, to set up the Lagrangian of the full theory yielding the equations of motion (\ref{eq:hameq}), and to treat \refeq{tcon} as a holonomic constraint to simplify the Lagrangian to an effective Lagrangian $L_{\rm eff}(\ff \eta, \dot{\ff \eta}, \ff m,\dot{\ff m})$ with a strongly reduced number of degrees of freedom.

In Appendix \ref{sec:lagtb} this program is carried out for an arbitrary function $\ff n_{0,i}(\ff \eta)$ without further specification.
The form of the resulting equation of motion, \refeq{ttimeseta}, 
\ba
0
=
\frac{\partial H_{\rm eff}(\ff \eta, \ff m)}{\partial \ff \eta} \times \ff \eta
+ 
\ff T \times \ff \eta
\: ,
\labeq{nodot}
\ea
turns out as quite unusual as it lacks an explicit $\dot{\ff \eta}$ term.
However, similar to the ASD, see \refeq{asd}, there is an additional topological spin-torque term resulting from the constraint. 
This has the form $\ff T = \dot{\ff \eta} \times \ff \Omega$, where $\ff \Omega$ is the pseudo-vector corresponding to an antisymmetric tensor $\Omega_{\mu\nu}$ that derives from the topological charge density \cite{EMP20} or magnetic vorticity \cite{Coo99}, see Eqs.\ (\ref{eq:omten1}) and (\ref{eq:omten2}).
This topological spin torque brings the $\dot{\ff \eta}$ dependency back into the theory.

Eq.\ (\ref{eq:nodot}) holds generally for constraints of the form $\ff n_{i} = \ff n_{0,i}(\ff \eta)$. 
In Appendix \ref{sec:eomtb}, we evaluate the topological charge density and the resulting topological spin torque for the constraint \refeq{tcon} explicitly.
This leads to equations of motion, which, for $\Delta = \sum_{i} z_{i} \ne 0$, have the familiar Hamiltonian form, cf.\ \refeq{eomtb}:
\ba
\Delta \dot{\ff \eta} 
&=&
S K \sum_{r} z_{i_{r}} \ff m_{r} \times \ff \eta
\: , 
\nonumber \\
\dot{\ff m}_{r}
&= &
s K z_{i_{r}} \ff \eta \times \ff m_{r} - S \ff B_{r} \times \ff m_{r} 
\: .
\labeq{eomcsd}
\ea
Some general properties and conservation laws related to these equations are discussed in Appendix \ref{sec:eomtb} as well.

\begin{figure}[t]
\includegraphics[width=0.95\columnwidth]{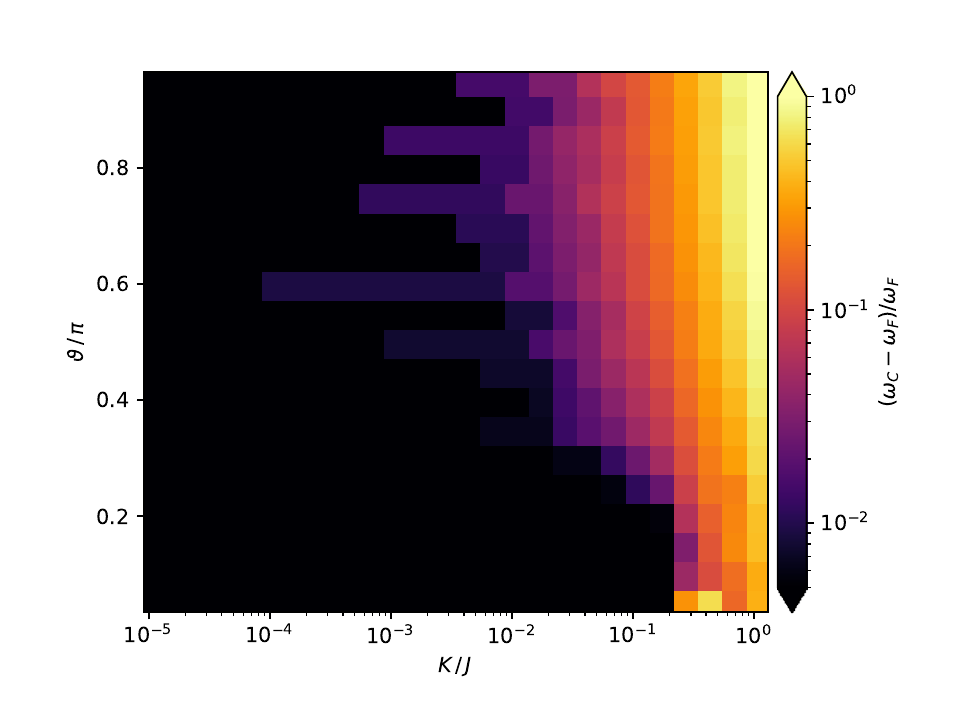}
\caption{
Main precession frequency in the Fourier spectrum of the real-time dynamics of $\ff m_{0}$ obtained from constrained spin-dynamics theory $\omega_{C}$ as function of $\vartheta$ and $K/J$.
Color code: normalized difference with the result of the full spin-dynamics theory $\omega_{F}$.
}
\label{fig:compc}
\end{figure}

\begin{figure}[b]
\includegraphics[width=0.95\columnwidth]{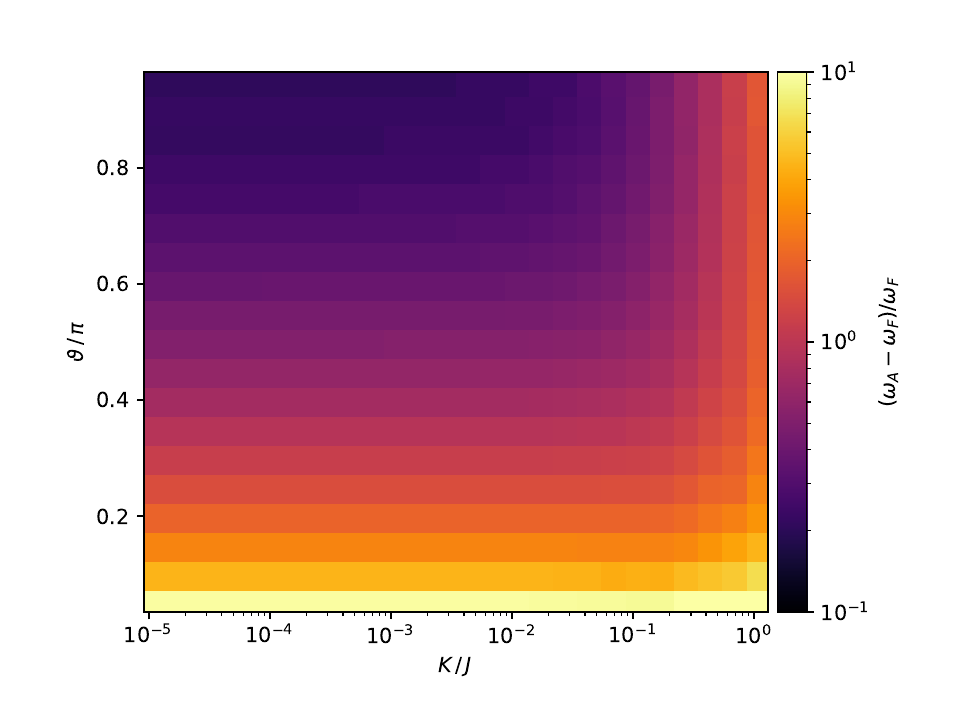}
\caption{
The same as Fig.\ \ref{fig:compc} but comparing the ASD and the full spin-dynamics theory.
}
\label{fig:compa}
\end{figure}

Here, we consider the setup discussed in the previous section, see Fig.\ \ref{fig:1d}, and compare the numerical solution of Eqs.\ (\ref{eq:eomcsd}) with that of the full set of equations of motion (\ref{eq:hameq}) and with the predictions of the ASD. 
For $R=2$ impurity spins, the constrained spin dynamics is in fact more complicated. 
In particular, there is no simple precessional motion at moderate $K/J$. 
For a comparison with the full spin-dynamics theory, \refeq{hameq}, we nevertheless concentrate on the dominant peak in the Fourier spectrum and the corresponding precession frequency $\omega_{\rm p}$.
Fig.\ \ref{fig:compc} demonstrates that spin dynamics under the tight-binding constraint in fact substantially improves the description and is reliable in the weak-coupling limit $K/J \ll 1$ {\em for all} angles $\vartheta$ specifying the initial impurity-spin configuration at time $t=0$.
This is opposed to the ASD, which requires weaker couplings $K/J$ and which captures the full spin dynamics for angles close to $\vartheta = \pi$ only, as is shown in Fig.\ \ref{fig:compa}.

The situation is completely different, however, for the case $\Delta=0$, i.e., if the chain of host spins consists of an even number of sites, such that for antiferromagnetic coupling $J$ the total host spin vanishes. 
Solving the equations of motion (\ref{eq:hameq}) of the full theory for $R=2$ impurity spins for $K/J\ll 1$, one finds a nontrivial spin dynamics with a precessional motion of $\ff \eta$ while $\ff m_{\rm tot}=\mbox{const}$.
For $R=1$, there is no spin dynamics at all, since for $\Delta=0$ the total spin is solely given by the single impurity spin, and, therefore, the impurity spin is fixed to its initial direction due to total spin conservation. 
Hence, we note that there are no special features here.

Turning to the constrained spin dynamics and specializing \refeq{eomcsd} to the case $\Delta=0$, $R=1$ and $\ff B=0$, provides us with the two equations $0 = \ff m \times \ff \eta$ and $\dot{\ff m} = s K \ff \eta \times \ff m$, which correctly imply $\ff m = \mbox{const}$. 
However, the first equation is dubious, since it may conflict with an initial condition where $\ff m \times \ff \eta \ne 0$.
On the other hand, such an initial state is perfectly allowed by our constraint \refeq{tcon}. 
This clearly implies that the constrained spin dynamics is inherently inconsistent. 

We have analyzed the origin of this inconsistency in Appendix \ref{sec:delta0}.
In fact, in the case $\Delta = 0$, the effective Lagrangian of the constrained spin-dynamics theory is singular. 
This can be made explicit with a proper gauge transformation after which $L_{\rm eff}(\ff \eta, \dot{\ff \eta}, \ff m,\dot{\ff m})$ becomes independent of $\dot{\ff \eta}$. 
Hence, it cannot describe situations where the $\ff \eta$ degrees of freedom are dynamic. 
Actually, this represents a clear example of a non-admissible effective Lagrangian theory. 

Let us finally turn to the case $\Delta\ne 0$ once more.
It is worth mentioning that the ASD can be newly derived by starting from the effective Lagrangian for spin dynamics under the tight-binding constraint $\ff n_{i} = \ff n_{0,i}(\ff \eta)$ and by imposing the {\em additional} constraint $\ff \eta = z_{K} \frac{\ff m_{0}}{m_{0}}$ expressing adiabaticity, see \refeq{eta}.
A heuristic argument is given in Appendix \ref{sec:single} for the single-impurity-spin case $R=1$. 
The formal derivation for the general case is worked out in Appendix \ref{sec:alt}.
This must be seen as a successful consistency check of the formal theory.

\section{Conclusions}
\label{sec:con}

Classical Heisenberg spin models are frequently used in atomistic spin-dynamics studies of condensed-matter systems, nanostructures or molecular systems. 
From a pragmatic point of view, they are quite attractive since the corresponding classical Hamiltonian equations of motion form a nonlinear set of ordinary differential equations, which can be integrated by numerical means, such that, as compared to quantum-spin models, long propagation times for a large number of spins are easily accessible. 
Usually, for generic model parameters, the resulting microscopic spin trajectories are chaotic and cover the entire accessible phase space, as it is expected for a nonlinear classical ergodic system. 

More regular dynamics is obtained for cases with strongly varying exchange-coupling parameters or, equivalently, for systems with a clear separation of intrinsic time scales. 
Such situations are often quite realistic, and a typical setup has been considered here.
We have performed a comprehensive study of a prototypical model consisting of two impurity spins that are weakly coupled to an antiferromagnetically coupled host-spin system, i.e., slow impurity spins are interacting with fast host spins. 
For initial states with energy close to the ground-state energy, a very regular, mainly precessional dynamics emerges, which calls for an effective low-energy theory.

The purely classical system studied here must be seen as a simple model system and would have to be refined to describe a realistic material that is accessible experimentally.
Several issues must be considered, such as longer-ranged interactions between the host spins, nonlocal coupling between impurity and host spins, anisotropies and more. 
Actually, the model studied here is probably the simplest one that serves our theoretical purposes.
The main conclusions, however, will all carry over qualitatively to more realistic setups.

Conceptually, the most interesting finding is that the spin dynamics is unexpectedly non-Hamiltonian in many cases, i.e., there is no effective RKKY-like Hamiltonian that merely consists of the impurity-spin degrees of freedom and is able to reproduce the impurity-spin dynamics. 
The reason is that, quite generally, the time-scale separation leads to the emergence of a topological spin torque, which profoundly affects the spin dynamics.
This is reminiscent of the Berry phase that emerges in a quantum (host) system upon slow variation of classical model parameters (the impurity spins). 
An important difference, however, is that the (purely classical) topological spin torque actually represents a {\em back-reaction} of the local topological charge density of the host system on the slow impurity spins.

Our main ansatz for constructing an effective low-energy impurity-spin dynamics has been the adiabatic approximation, which is formulated as a constraint for the host-spin configuration. 
This constraint has to be incorporated carefully:
Making use of the constraint on the level of the equations of motion runs into unacceptable inconsistencies. 
A consistent effective theory is obtained when using the constraint to simplify the original Hamiltonian. 
This naive effective theory, however, runs the risk of not respecting certain conservation laws, e.g., total spin conservation and has been explicitly shown to fail in cases, where the host-spin system has a finite total spin moment. 
A satisfactory effective theory rather requires to work in the Lagrange formalism which allows us to include arbitrary constraints in a consistent way. 
Using the adiabatic constraint defines adiabatic spin dynamics (ASD).
For the relevant weak-coupling limit, we were able to work out the non-Hamiltonian effective equations of motion analytically. 
The big impact of the topological spin torque appearing in the ASD equations becomes evident when comparing the ASD results with those of the naive theory. 

From a theoretical perspective, the ASD appears as a very attractive approach: 
It follows a clear construction principle, it maintains conservation laws resulting from the symmetries of the original Hamiltonian, it provides a true effective theory formulated in terms of the slow impurity-spin degrees of freedom only, and it brings a hidden topological structure to light that substantially modifies the slow spin dynamics.
On the other hand, the applicability of the ASD stands and falls with the validity of the constraint imposed, and unfortunately, contrary to quantum systems, there is no direct classical equivalent of the adiabatic theorem which ensures adiabaticity in certain limits.
Comparison of the predictions of the ASD with those of the full theory treating all, slow and fast degrees of freedom, is thus necessary. 
In fact, this has uncovered some deficiencies:
While the ASD applies to the weak-coupling limit only, as it was anticipated, it also requires that the initial impurity-spin configuration is not too far from the ground-state configuration, and even in this case there is a good but not fully convincing agreement with the full theory.

We have therefore studied another version of a constrained spin dynamics assuming that the host-spin system is tightly bound but not necessarily in the ground state for the present impurity-spin configuration at any instant of time.
Also this constrained spin dynamics must be worked out carefully within the Lagrange formalism, and again there is a topological spin torque involved.
Spin dynamics under the tight-binding constraint somewhat relaxes the ASD constraint. 
In fact, the ASD could be newly derived by enforcing the missing piece again. 
Comparing with the full theory, we found that the relaxation of the constraint indeed results in an improved effective theory, which now covers the entire weak-coupling limit. 
This advantage, however, also comes at a cost: 
Spin dynamics under the tight-binding constraint necessarily involves host degrees of freedom, i.e., it fails to provide an effective impurity-spin dynamics theory. 
More severely, however, the effective Lagrangian is singular in the case of a nonmagnetic host with a vanishing total spin.

Our present study can be seen as a first step towards an effective theory of RKKY real-time dynamics, i.e., where impurity spins are coupled to a conduction-electron system, and work on this quantum-classical problem in already in progress.
Clearly, this problem is more involved since with the Fermi energy of the electronic system there is an 
additional energy scale to be considered. 
This also implies the emergence of a length scale, resulting, e.g., in the nontrivial distance dependence of the effective RKKY exchange. 
Furthermore, we expect the resulting effective theory to be of non-Hamiltonian character as well. 
We also expect to make contact with the Berry curvature of the electronic system and a corresponding topological spin torque, replacing the topological charge density of the purely classical host-spin system studied here. 
Clearly, the quantum-classical problem is more relevant for interpreting experimental findings. 
We believe that the insights gained from our present classical study will be very helpful for this next step.

\acknowledgments
This work was supported by the Deutsche Forschungsgemeinschaft (DFG) through the Cluster of Excellence ``Advanced Imaging of Matter'' - EXC 2056 - project ID 390715994, and by the DFG 
Sonderforschungsbereich 925 ``Light-induced dynamics and control of correlated quantum systems''
(project B5).

\appendix

\begin{widetext}

\section{Total spin conservation within the ASD}
\label{sec:spincons}

Within adiabatic spin dynamics the total spin, i.e., the sum of the total impurity spin $\ff S_{\rm tot}$ and the total host spin $\ff s_{\rm tot}$, is conserved, if $H_{\rm eff}$ is SO(3) symmetric. 
Here, we prove total-spin conservation for a collinear host-spin structure. 
We start by computing the time derivative of the total spin:
\be
\frac{d}{dt} ( \ff S_{\rm tot}+ \ff s_{\rm tot} )
=
S \sum_{r} \dot{\ff m}_{r} + s \sum_{i} \frac{d}{dt}  \ff n_{0,i}(\ff m)
=
\sum_{r} \frac{\partial H_{\rm eff}}{\partial \ff m_{r}} \times \ff m_{r} 
+ 
\sum_{r} \ff T_{r} \times \ff m_{r}
+ 
s \sum_{ir\mu} \dot{m}_{r\mu} \frac{\partial \ff n_{0,i}(\ff m)}{\partial m_{r\mu}}  \: ,
\ee
where, in the first step, we have inserted the equation of motion \refeq{asd} to eliminate $\dot{\ff m}_{r}$. 
Consider the second term in the last expression. 
Making use of \refeq{sumtm} we find:
\be
\sum_{r} \ff T_{r} \times \ff m_{r}
=
\sum_{r} z_{i_{r}} z_{K} \frac{s\Delta}{m^{3}_{0}} (\dot{\ff m}_{0} \times \ff m_{0}) \times \ff m_{r}  
=
z_{K} \frac{s\Delta}{m^{3}_{0}} (\dot{\ff m}_{0} \times \ff m_{0}) \times \ff m_{0} \: . 
\ee
To treat the third term, we employ \refeq{dn0dm}:
\ba
s \sum_{ir\mu} \dot{m}_{r\mu} \frac{\partial \ff n_{0,i}(\ff m)}{\partial m_{r\mu}}  
&=&
s \sum_{ir\mu} z_{i} z_{K} \dot{m}_{r\mu} z_{i_{r}} \frac{1}{m_{0}} 
\left( \ff e_{\mu} - \frac{m_{0\mu}}{m_{0}^{2} } \ff m_{0} \right)
\nonumber \\
&=&
s z_{K} \sum_{i} z_{i} \frac{1}{m_{0}} 
\left( 
\dot{\ff m}_{0} - (\dot{\ff m}_{0} \ff m_{0}) \frac{\ff m_{0}}{m_{0}^{2} }
\right)
=
z_{K} s \Delta \frac{1}{m_{0}^{3}} 
\ff m_{0} \times ( \dot{\ff m}_{0} \times \ff m_{0}) \: .
\ea
This cancels the second term.
SO(3) symmetry implies that $H_{\rm eff}(\ff m)$ has the general form
\be
  H_{\rm eff}(\ff m) = f((c_{rr'})_{r,r'=1,...,R}) \: , 
\ee
where $f$ is an arbitrary smooth function of all inner products $c_{rr'} \equiv \ff m_{r} \ff m_{r'}$.
Since $\ff m_{r}^{2}=1$, we have
\be
  \sum_{r} \frac{\partial H_{\rm eff}}{\partial \ff m_{r}} \times \ff m_{r}
  =
  \sum_{r} \sum_{r'r''} \frac{\partial f}{\partial c_{r'r''}} \frac{\partial{c_{r'r''}}}{\partial \ff m_{r}} \times \ff m_{r}
  =
  \sum_{r} \sum_{r'r''} \frac{\partial f}{\partial c_{r'r''}} \left( \delta_{rr'} \ff m_{r''} + \delta_{rr''} \ff m_{r'} \right)
   \times \ff m_{r}
   = 0 \: .
\ee
This proves that $\ff S_{\rm tot}+ \ff s_{\rm tot} = \mbox{const}$.

\end{widetext}

\section{Solution of coupled ODE's}
\label{sec:ode}

We consider the following system of two (three-component) ordinary differential equations
\ba
  \dot{\ff x}_{1} &=& c_{1} \ff x_{2} \times \ff x_{1} \; ,
  \nonumber \\
  \dot{\ff x}_{2} &=& c_{2} \ff x_{1} \times \ff x_{2} \; ,
\labeq{o4}  
\ea
where $c_{1}, c_{2}$ are constants.
With the scaling transformation 
\be
  \ff y_{1} = \alpha_{1} \ff x_{1} \; , \quad
  \ff y_{2} = \alpha_{2} \ff x_{2} \; ,
\ee
where $\alpha_{1}, \alpha_{2}$ are constants to be determined, we have
\ba
  \dot{\ff y}_{1} &=& \frac{c_{1}}{\alpha_{2}} \ff y_{2} \times \ff y_{1} \; ,
  \nonumber \\
  \dot{\ff y}_{2} &=& \frac{c_{2}}{\alpha_{1}} \ff y_{1} \times \ff y_{2} \; .
\ea
Choosing
\be
  \alpha_{1} = \sqrt{\left| \frac{c_{2}}{c_{1}} \right|} \; , \quad
  \alpha_{2} = \sqrt{\left| \frac{c_{1}}{c_{2}} \right|} = \frac{1}{\alpha_{1}} \; , 
\labeq{o3}  
\ee
we get
\ba
  \dot{\ff y}_{1} &=& s_{1} \sqrt{|c_{1}c_{2}|} \, \ff y_{2} \times \ff y_{1} \; ,
  \nonumber \\
  \dot{\ff y}_{2} &=& s_{2} \sqrt{|c_{1}c_{2}|} \, \ff y_{1} \times \ff y_{2} \; ,
\labeq{o1}
\ea
where $s_{1} = c_{1} / |c_{1}|$ and $s_{2} = c_{2} / |c_{2}|$ are sign factors. 
We distinguish the two cases $s_{1} = \pm s_{2}$ and conclude that there is a conserved vector
\be
  \ff y
  \equiv
  \ff y_{1} \pm \ff y_{2} 
  = 
  \sqrt{\left| \frac{c_{2}}{c_{1}} \right|} \ff x_{1}
  \pm 
  \sqrt{\left| \frac{c_{1}}{c_{2}} \right|} \ff x_{2}
  = \mbox{const.}
\labeq{o2}
\ee
Since Eqs.\ (\ref{eq:o1}) and (\ref{eq:o2}) imply
\ba
  \dot{\ff y}_{1} &=& \pm s_{1} \sqrt{|c_{1}c_{2}|} \, \ff y \times \ff y_{1} \; ,
  \nonumber \\
  \dot{\ff y}_{2} &=& s_{2} \sqrt{|c_{1}c_{2}|} \, \ff y \times \ff y_{2} \; ,
\ea
we see that $\ff y_{1}$ and $\ff y_{2}$ and, thus, $\ff x_{1}$ and $\ff x_{2}$ precess with equal orientation around $\ff y$. 
To compute the precession frequency $\omega_{\rm p} = \sqrt{|c_{1}c_{2}|} \, |\ff y|$ we need the length of $\ff y$:
\ba
  |\ff y|^{2} 
  &=& 
  |\ff y_{1}|^{2} + |\ff y_{2}|^{2} \pm 2 \ff y_{1} \ff y_{2} 
  \nonumber \\
  &=&
  \left| \frac{c_{2}}{c_{1}} \right| x_{1}^{2} + \left| \frac{c_{1}}{c_{2}} \right| x_{2}^{2} \pm 2 \ff x_{1} \ff x_{2}
\: .
\ea
Here, we have used \refeq{o3} and $\alpha_{1} \alpha_{2} = 1$ in particular.
Note that \refeq{o4} immediately implies that $x_{1}, x_{2}$ and $\ff x_{1} \ff x_{2}$ are constant.
The precession frequency
\be
  \omega_{\rm p}
  =
  \sqrt{
  c_{2}^{2} x_{1}^{2} 
  + 
  c_{1}^{2} x_{2}^{2} 
  \pm 2 \left| c_{1} c_{2} \right| \, \ff x_{1} \ff x_{2}
  }
\ee
depends on the lengths of $\ff x_{1}$ and $\ff x_{2}$ and on the angle enclosed initially.

Let us now consider the differential equations \refeq{msexpl} and \refeq{mtdyn} for $\ff x_{1} = \ff m_{0}$ and $\ff x_{2} = \ff m_{\rm tot}$.
The coefficients corresponding to $\ff m_{0}$ and $\ff m_{\rm tot}$ read
\ba
c_{0} & = &
\frac{1}{
1+z_{K} \frac{s\Delta}{S m^{3}_{0}} \ff m_{0} \ff m_{\rm tot}
}
\frac{|K|  s}{m_{0}} 
\;  ,
\nonumber \\
c_{\rm tot} & = &
 z_{K} \frac{s\Delta}{S m_{0}}  
\frac{1}{
1+z_{K} \frac{s\Delta}{S m^{3}_{0}} \ff m_{0} \ff m_{\rm tot}
}
\frac{|K|  s}{m_{0}} \; , 
\ea
respectively. 
With 
$c_{\rm tot} / c_{0}
=
z_{K} s \Delta / S m_{0}$, 
this means precession around the axis 
\be
  \ff y
  = 
  \sqrt{ \frac{s}{S} \frac{|\Delta|}{m_{0}} } \, \ff m_{0}
  \pm 
  \sqrt{ \frac{S}{s} \frac{m_{0}}{|\Delta|} } \, \ff m_{\rm tot}
  = \mbox{const.}
\ee
where the ``$+$''-sign applies for $z_{K}\Delta >0$ and the ``$-$''-sign for $z_{K}\Delta <0$.
After rescaling, we note that $\ff y$ is collinear to 
\be
  \pm \frac{s |\Delta|}{m_{0}}  \, \ff m_{0}
  +
  S \ff m_{\rm tot}
  = \mbox{const.}
\labeq{totot}
\ee
Since $\ff s_{\rm tot} = s \sum_{i} z_{i} \ff \eta = s \Delta z_{K} \ff m_{0} / m_{0} = \pm s |\Delta| \ff m_{0} / m_{0}$, this means that the precession axis is just defined by the total spin $\ff s_{\rm tot} + \ff S_{\rm tot}$, as expected on physical grounds.

A simple result for the precession frequency is obtained in the case of two impurity spins ($R=2$) with $z_{i_{1}} = - z_{i_{2}}$, where we can exploit the relation $\ff m_{0} \ff m_{\rm tot} = (\ff m_{1} - \ff m_{2}) (\ff m_{1} + \ff m_{2}) = 0$:
\be
  \omega_{\rm p} = |K| s \, \frac{1}{m_{0}} 
  \sqrt{
  \frac{s^2 |\Delta|^2}{S^2} 
  +
  m_{\rm tot}^{2}
  } \: .
\ee
Assuming that $s=S$ and that $|\Delta|=1$ (antiferromagnetic host-spin configuration and odd $L$), 
the precession frequency is
\be
  \omega_{\rm p} 
  =
  |K| s\, \frac{\sqrt{1 + m_{\rm tot}^{2}}}{m_{0}}
   =
  |K| s \, \frac{1}{2\sin\nicefrac{\vartheta}2}
 \sqrt{1 + 4  \cos^2 \frac{\vartheta}2}
  \: ,
  \labeq{pfapp}  
\ee
where $\vartheta \in \, ] 0,\pi ]$ is the conserved angle enclosed by $\ff m_{1}$ and $\ff m_{2}$.

\section{Using tight-binding constraints to simplify the equations of motion}
\label{sec:tb}

In an attempt to construct an alternative effective theory, let us start from the fundamental equations of motion \refe{hameq} for the impurity spins $\ff S_{r} = S \ff m_{r}$.
Using the notations of Sec.\ \ref{sec:mod}, we have
\be
\dot{\ff m}_{r} = K s \, \ff n_{i_{r}} \times \ff m_{r} \; ,
\ee
where we have assumed $\ff B_{r}=0$, for simplicity. 
Further, the equations of motion for $\ff s_{i_{r}} = s \ff n_{i_{r}}$ read
\be
\dot{\ff n}_{i_{r}} = K S \,\ff m_{r} \times \ff n_{i_{r}} + s \sum_{i'} J_{i_{r} i'} \ff n_{i'} \times \ff n_{i_{r}}
\: .
\labeq{nais}
\ee
We want to exploit the constraint 
\be
\ff n_{i} = z_{i} \ff \eta
\labeq{con1}
\ee 
($z_{i}= \pm 1$). 
This tight-binding constraint expresses that for $K \ll J$ all host spins are tightly bound together such that, irrespective of the impurity-spin configuration, all $\ff n_{i}$ are collinear to a unit vector $\ff \eta$ at all times $t$.

It is tempting, but incorrect, to use the constraint to simplify the equations of motion as will be shown here.
\refeq{con1} implies that the second term in \refeq{nais} vanishes.
Using the constraint once more, we can eliminate the host spins and are left with 
\be
\dot{\ff m}_{r} = K s z_{i_{r}} \ff \eta \times \ff m_{r} \; , 
\ee
and 
\be
\dot{\ff \eta} = K S \, \ff m_{r} \times \ff \eta 
\ee
for all $r=1,..., R$.
This yields
\be
(\ff m_{r} - \ff m_{r'}) \times \ff \eta= 0
\: , 
\ee
or
\be
 \ff \eta = \pm \frac{\ff m_{r} - \ff m_{r'}}{|\ff m_{r} - \ff m_{r'}|}
\: , 
\ee
for all $r,r'$. 
For arbitrary directions $\ff m_{r}$ and for $R\ge 3$, however, this obviously leads to contradictions.

\begin{widetext}

\section{Lagrange formalism using tight-binding constraints}
\label{sec:lagtb}

The correct dynamics under a constraint of the form $\ff n = \ff n_{0}(\ff \eta)$ ($\ff n_{i} = \ff n_{0,i}(\ff \eta)$) can be derived from the effective Lagrangian 
\be
L_{\rm eff}(\ff \eta, \dot{\ff \eta}, \ff m,\dot{\ff m}) \equiv L(\ff n_{0}(\ff \eta),(d/dt)\ff n_{0}(\ff \eta),\ff m,\dot{\ff m}) \: ,
\label{eq:leff0}
\ee
where $L(\ff n,\dot{\ff n},\ff m,\dot{\ff m}) = S \sum_{j} \ff A(\ff m_{j}) \dot{\ff m}_{j} + s \sum_{i} \ff A(\ff n_{i}) \dot{\ff n_{i}} - H(\ff n, \ff m)$ is the full Lagrangian.
Here, we use the short-hand notation $\ff n_{0}=(\ff n_{0,1}, ..., \ff n_{0,L})$,  $\ff n=(\ff n_{1}, ..., \ff n_{L})$ and $\ff m =(\ff m_{1}, ..., \ff m_{R})$. 
Furthermore, $\ff A(\ff r)$ is a vector field satisfying $\nab \times \ff A(\ff r)=-\ff r / r^{3}$, and which can thus be interpreted as the vector potential of a unit magnetic (Dirac) monopole located at $\ff r = 0$. 
In the standard gauge \cite{Dir31}, this is given by $\ff A(\ff r) = - (1/r^{2}) (\ff e_{z} \times \ff r) / (1 + \ff e_{z} \ff r / r)$. 
The equations of motion deriving from the full Lagrangian are equivalent with the Hamilton equations (\ref{eq:hameq}), see Ref.\ \cite{EMP20} for further details.

With
\be
\frac{d}{dt} \ff n_{0,i}(\ff \eta) = (\dot{\ff \eta} \nab) \ff n_{0,i}(\ff \eta)
\ee
we find:
\ba
L_{\rm eff}(\ff \eta, \dot{\ff \eta}, \ff m,\dot{\ff m}) 
=
S \sum_{j} \ff A(\ff m_{j}) \dot{\ff m}_{j}
+
s \sum_{i} \ff A(\ff n_{0,i}(\ff \eta)) 
\Big((\dot{\ff \eta} \nab) \ff n_{0,i}(\ff \eta) \Big)
-
H_{\rm eff}(\ff \eta, \ff m) \; , 
\label{eq:leff}
\ea
where $i=1,...,L$ and $j=1,...,R$, and where $H_{\rm eff}(\ff \eta, \ff m) = H(s \ff n_{0}(\ff \eta),S \ff m)$.
To get the Lagrange equations of motion, we first compute
\ba
\frac{\partial L_{\rm eff}(\ff \eta, \dot{\ff \eta}, \ff m, \dot{\ff m})}{\partial \ff m_{r}} 
&=&
S \sum_{\beta}
\nab_{r} A_{\beta}(\ff m_{r}) \dot{m}_{r\beta}
-
\frac{\partial H_{\rm eff}(\ff \eta, \ff m)}{\partial \ff m_{r}}  \; .
\label{eq:dldm1}
\ea
and
\ba
\frac{\partial L_{\rm eff}(\ff \eta, \dot{\ff \eta}, \ff m, \dot{\ff m})}{\partial \ff \eta} 
=
s \sum_{i\beta} A_{\beta}(\ff n_{0,i}(\ff \eta))
\: (\dot{\ff \eta} \nab) \nab n_{0,i\beta}(\ff \eta)
+
s \sum_{i\alpha\beta} 
\frac{\partial A_{\beta}(\ff n_{0,i}(\ff \eta))}{\partial n_{0,i\alpha}}
\nab n_{0,i\alpha}(\ff \eta)
(\dot{\ff \eta} \nab)  n_{0,i\beta}(\ff \eta)
-
\frac{\partial H_{\rm eff}(\ff \eta, \ff m)}{\partial \ff \eta}  \; .
\nonumber \\
\label{eq:dldm2}
\ea
Here, $\nab_{r} = \partial / \partial \ff m_{r}$, and Greek indices $\alpha, \beta, ... \in \{x,y,z\}$. 
Furthermore, 
\ba
\frac{\partial L_{\rm eff}(\ff \eta, \dot{\ff \eta}, \ff m,\dot{\ff m})}{\partial \dot{\ff m}_{r}} 
=
S \ff A(\ff m_{r}) 
\; ,
\quad
\frac{\partial L_{\rm eff}(\ff \eta, \dot{\ff \eta}, \ff m,\dot{\ff m})}{\partial \dot{\ff \eta}} 
=
s \sum_{i\alpha} A_{\alpha}(\ff n_{0,i}(\ff \eta))   
\nab n_{0,i\alpha}(\ff \eta) 
\; ,
\label{eq:deta}
\ea
which yields
\ba
\frac{d}{dt} \frac{\partial L_{\rm eff}(\ff \eta, \dot{\ff \eta}, \ff m, \dot{\ff m})}{\partial \dot{\ff m}_{r}} 
&=&
S
(\dot{\ff m}_{r} \nab_{r}) \ff A(\ff m_{r}) 
\ea
and
\ba
\frac{d}{dt} \frac{\partial L_{\rm eff}(\ff \eta, \dot{\ff \eta}, \ff m, \dot{\ff m})}{\partial \dot{\ff \eta}} 
&=&
s \sum_{i\alpha\beta} \frac{\partial A_{\alpha}(\ff n_{0,i}(\ff \eta))}{\partial n_{0,i\beta}}
(\dot{\ff \eta} \nab n_{0,i\beta}(\ff \eta) )
\nab n_{0,i\alpha}(\ff \eta)
+
s \sum_{i\alpha} A_{\alpha}(\ff n_{0,i}(\ff \eta) 
\nab
\left(
\dot{\ff \eta} \nab n_{0,i\alpha}(\ff \eta) 
\right)
\; .
\ea
The last term equals the first term on the right-hand side of Eq.\ (\ref{eq:dldm2}) in the Lagrange equations, since $\nab$ and $\dot{\ff \eta} \nab$ commute, such that we are left with:
\ba
0 &=& \frac{d}{dt} \frac{\partial L_{\rm eff}}{\partial \dot{\ff m}_{r}} 
-
\frac{\partial L_{\rm eff}}{\partial {\ff m}_{r}} 
=
S
(\dot{\ff m}_{r} \nab_{r}) \ff A(\ff m_{r}) 
-
S \sum_{\beta}
\nab_{r} A_{\beta}(\ff m_{r}) \dot{m}_{r\beta}
+
\frac{\partial H_{\rm eff}(\ff \eta, \ff m)}{\partial \ff m_{r}} 
\nonumber \\
&=&
S (\nab_{r} \times \ff A(\ff m_{r})) \times \dot{\ff m}_{r} 
+
\frac{\partial H_{\rm eff}(\ff \eta, \ff m)}{\partial \ff m_{r}} 
\label{eq:eomtb1}
\: ,
\ea
and
\ba
0 &=& \frac{d}{dt} \frac{\partial L_{\rm eff}}{\partial \dot{\ff \eta}} 
-
\frac{\partial L_{\rm eff}}{\partial {\ff \eta}} 
=
\frac{\partial}{\partial \ff \eta} H_{\rm eff}(\ff \eta, \ff m) 
-
s \sum_{i\alpha\beta} 
\frac{\partial A_{\beta}(\ff n_{0,i}(\ff \eta))}{\partial n_{0,i\alpha}}
\nab n_{0,i\alpha}(\ff \eta)
(\dot{\ff \eta} \nab)  n_{0,i\beta}(\ff \eta)
\nonumber \\
&+& 
s \sum_{i\alpha\beta} \frac{\partial A_{\alpha}(\ff n_{0,i}(\ff \eta))}{\partial n_{0,i\beta}}
(\dot{\ff \eta} \nab n_{0,i\beta}(\ff \eta) )
\nab n_{0,i\alpha}(\ff \eta)
\nonumber \\
&=&
\frac{\partial H_{\rm eff}(\ff \eta, \ff m)}{\partial \ff \eta} 
+ \ff T
\label{eq:eomtb2}
\: ,
\ea
where $\ff T$ stands for the last two terms. 
Taking in \refeq{eomtb1} the cross product from the right, $(...)\times \ff m_{r}$, we find
\ba
S ((\nab_{r} \times \ff A(\ff m_{r})) \times \dot{\ff m}_{r} ) \times \ff m_{r} 
+
\frac{\partial H_{\rm eff}(\ff \eta, \ff m)}{\partial \ff m_{r}} \times \ff m_{r}
= 0
\: .
\label{eq:eomtb3}
\ea
Using $\nab_{r} \times \ff A(\ff m_{r}) = - \ff m_{r} / m_{r}^{3}$, expanding the remaining double cross product and exploiting that $\ff m_{r}$ is a unit vector, yields:
\be
S \dot{\ff m}_{r}
=
\frac{\partial H_{\rm eff}(\ff \eta, \ff m)}{\partial \ff m_{r}} \times \ff m_{r}
\: , 
\ee
which just recovers the standard form of the equation of motion for $\ff m_{r}$.
On the contrary, the equation of motion for $\ff \eta$, which is obtained from \refeq{eomtb2} by taking the cross product with $\ff \eta$, is unconventional:
\ba
0
=
\frac{\partial H_{\rm eff}(\ff \eta, \ff m)}{\partial \ff \eta} \times \ff \eta
+ 
\ff T \times \ff \eta
\: .
\labeq{ttimeseta}
\ea

\end{widetext}

Note that actually we should have added Lagrange-multiplier terms, $L_{\rm eff}(\ff \eta, \dot{\ff \eta}, \ff m, \dot{\ff m}) \mapsto L_{\rm eff}(\ff \eta, \dot{\ff \eta}, \ff m, \dot{\ff m}) -\sum_{r} \lambda_{r} (\ff m_{r}^{2}-1) - \lambda (\ff \eta^{2}-1)$, to account for the normalization conditions $\ff m_{r}^{2}=1$ and $\ff \eta^{2}=1$.
However, this would merely have resulted in additional summands $2\lambda_{r} \ff m_{r}$ and  $2\lambda \ff \eta$ on the right-hand sides of Eqs.\ (\ref{eq:eomtb1}) and (\ref{eq:eomtb2}), respectively, which do not contribute after taking the respective cross products $(...) \times \ff m_{r}$ and $(...)\times \ff \eta$.
On the other hand, taking the dot products, $(...)\cdot \ff m_{r}$ and $(...)\cdot \ff \eta$, in Eqs.\ (\ref{eq:eomtb1}) and (\ref{eq:eomtb2}), respectively, just yields the necessary conditional equations for $\lambda_{r}$ and $\lambda$, if these were required.

$\ff T$ gives rise to a geometrical spin torque $\ff T \times \ff \eta$ and can be read off from Eq.\ (\ref{eq:eomtb2}):
\ba
\ff T
&=&
s
\sum_{i\alpha\beta} 
\left( 
\frac{\partial A_{\alpha}(\ff n_{0,i}(\ff \eta))}{\partial n_{0,i\beta}} 
-
\frac{\partial A_{\beta}(\ff n_{0,i}(\ff \eta))}{\partial n_{0,i\alpha}}
\right)
\nonumber \\ 
&\cdot&
(\dot{\ff \eta} \nab) n_{0,i\beta}(\ff \eta) 
\,
\nab n_{0,i\alpha}(\ff \eta) \: .
\ea
Exploiting once more the defining property of the vector potential, $\nab_{i} \times \ff A(\ff n_{0,i}) = - \ff n_{0,i} / n_{0,i}^{3}$, and using the normalization $n_{0,i}=1$ in the end, we find:
\ba
\ff T
&=&
s \sum_{i\alpha\beta\gamma}
\epsilon_{\alpha\beta\gamma}
\nab n_{0,i\alpha}(\ff \eta) \,
(\dot{\ff \eta} \nab) n_{0,i\beta}(\ff \eta) \, 
n_{0,i\gamma}(\ff \eta) 
\nonumber \\
&=&
s \sum_{i} \sum_{\mu\nu}
\nabla_{\mu} \ff n_{0,i}(\ff \eta) 
\times
\nabla_{\nu}   \ff n_{0,i}(\ff \eta) 
\cdot
\ff n_{0,i}(\ff \eta) 
\:\dot{\eta}_{\nu} \, \ff e_{\mu} \: . 
\nonumber \\
\ea
The scalar triple product defines an antisymmetric tensor of rank two:
\ba
\Omega_{\mu\nu}
&=&
s \sum_{i} 
\frac{\partial \ff n_{0,i}(\ff \eta) }{\partial \eta_{\mu}}
\times
\frac{\partial \ff n_{0,i}(\ff \eta) }{\partial \eta_{\nu}}
\cdot
\ff n_{0,i}(\ff \eta) 
\nonumber \\
&=&
- \Omega_{\nu\mu} 
= 
\sum_{\rho}
\epsilon_{\mu\nu\rho} \Omega_{\rho}
\: ,
\labeq{omten1}
\ea
where the last equation defines the pseudovector $\ff \Omega$ with components $\Omega_{\rho} = \frac12 \sum_{\mu\nu} \epsilon_{\mu\nu\rho} \Omega_{\mu\nu}$:
\be
\ff \Omega 
=
\frac{s}{2} \sum_{i} \sum_{\alpha\beta\gamma} \epsilon_{\alpha\beta\gamma} 
\nab n_{0,i\alpha} \times \nab n_{0,i\beta} \, n_{0,i\gamma}
\; , 
\labeq{omten2}
\ee
which has precisely the form of the ``magnetic vorticity'' \cite{Coo99}.
Hence:
\be
   \ff T
   =
   \sum_{\mu\nu}
   \Omega_{\mu\nu}
   \:
   \dot{\eta}_{\nu} \, \ff e_{\mu}
   =
   \dot{\ff \eta} \times \ff \Omega
   \: .
\labeq{eomlagtb}
\ee
Note that $\ff T \dot{\ff \eta} = 0$. 
Inserting the result for $\ff T$ in the equation of motion, we obtain 
\ba
0
=
\frac{\partial H_{\rm eff}(\ff \eta, \ff m)}{\partial \ff \eta} \times \ff \eta
+ 
(\dot{\ff \eta} \times \ff \Omega) \times \ff \eta
\: .
\ea
If the pseudovector $\ff \Omega$ is interpreted as a magnetic field in $\ff \eta$-space, $\ff T=\dot{\ff \eta} \times \ff \Omega$ is the Lorentz force (per unit charge) and $\ff T \times \ff \eta$ the corresponding torque.
On the other hand, the analogy cannot be made complete, as the curl of the vector potential of a ``magnetic monopole'', $\nab_{i} \times \ff A(\ff n_{0,i}) = - \ff n_{0,i} / n_{0,i}^{3}$, is a field in $\ff n_{0,i}$-space.

\section{Spin dynamics under tight-binding constraints}
\label{sec:eomtb}

Starting from the constraint, $\ff n_{i} = \ff n_{0,i}(\ff \eta) = z_{i} \ff \eta$, the computation of the topological spin torque is straightforward. 
We have:
\be
\frac{\partial \ff n_{0,i}(\ff \eta) }{\partial \eta_{\mu}}
=
z_{i} \ff e_{\mu}
\ee
and thus
\be
\Omega_{\mu\nu}
=
s \sum_{i} 
(z_{i} \ff e_{\mu})
\times
(z_{i} \ff e_{\nu})
\cdot
(z_{i} \ff \eta)
=
s \ff \eta \sum_{i} z_{i}
\ff e_{\mu} \times \ff e_{\nu}
\; , 
\ee
or in terms of the psuedovector
\be
\ff \Omega
= s \Delta \, \ff \eta
\: ,
\ee
with 
\be
\Delta \equiv \sum_{i=1}^{L} z_{i} \: .
\ee
For an antiferromagnetic host, e.g., we have $\Delta = \pm 1$ if $L$ is odd, and $\Delta = 0$ if $L$ is even.
Generally, $\ff \Omega$ is just the total host spin:
\be
\ff s_{\rm tot} 
= 
\sum_{i} \ff s_{i} 
= 
\sum_{i} s z_{i} \ff \eta 
= 
s \Delta \, \ff \eta
= 
\ff \Omega \: .
\ee

Now, the topological spin torque reads as
\ba
  \ff T \times \ff \eta 
  &=& 
  (\dot{\ff \eta} \times \ff \Omega) \times \ff \eta
  =
  s \Delta \, (\dot{\ff \eta} \times \ff \eta) \times \ff \eta
\nonumber \\
  &=& - s \Delta \, \dot{\ff \eta} 
  =
 - s \dot{\ff n}_{\rm tot}
 =
 - \dot{\ff s}_{\rm tot} \: , 
\ea
and therefore the set of equations of motion is given by
\ba
s \Delta \dot{\ff \eta}
&=&
\frac{\partial H_{\rm eff}(\ff \eta, \ff m)}{\partial \ff \eta} \times \ff \eta
\: , 
\nonumber \\
S \dot{\ff m}_{r}
&=&
\frac{\partial H_{\rm eff}(\ff \eta, \ff m)}{\partial \ff m_{r}} \times \ff m_{r} \: .
\ea
We see that, for $\Delta \ne 0$ and opposed to the ASD, one arrives at a standard Hamiltonian dynamics for the remaining degrees of freedom $\ff \eta$ and $\ff m$ governed by an effective Hamiltonian which is obtained by making use of the constraint in the original Hamiltonian.

Explicitly, the effective Hamiltonian, $H_{\rm eff}(\ff \eta, \ff m) = H(\ff s_{i} = s z_{i} \ff \eta , \ff S_{r} = S \ff m_{r})$, reads 
\be
H_{\rm eff}(\ff \eta, \ff m) 
=
E_{0}
+
K s S \ff \eta \sum_{r=1}^{R}  z_{i_{r}} \ff m_{r}
-
S \sum_{r=1}^{R} \ff m_{r} \ff B_{r} \: .
\labeq{effham}
\ee
This describes a central spin model: 
The impurity spins $\ff S_{r} = S \ff m_{r}$ couple with strengths $z_{i_{r}} K$ to the central spin $\ff s_{\rm tot} = s \ff n_{\rm tot} = s \Delta \ff \eta$.
With 
\ba
  \frac{\partial H_{\rm eff}(\ff \eta, \ff m)}{\partial \ff \eta} 
  &=&
  s S K \sum_{r} z_{i_{r}} \ff m_{r}
  \; , \nonumber \\
  \frac{\partial H_{\rm eff}(\ff \eta, \ff m)}{\partial \ff m_{r}} 
  &=&
  s S K z_{i_{r}} \ff \eta - S \ff B_{r}
  \: ,
\labeq{dheff}  
\ea
the Hamiltonian equations are given by:
\ba
s \Delta \dot{\ff \eta} 
&=&
s S K \sum_{r} z_{i_{r}} \ff m_{r} \times \ff \eta
\: , 
\nonumber \\
S \dot{\ff m}_{r}
&= &
s S K z_{i_{r}} \ff \eta \times \ff m_{r} - S \ff B_{r} \times \ff m_{r} 
\: .
\labeq{eomtb}
\ea

Let us start the discussion with the case $\Delta \ne 0$ and derive some consequences of the equations of motion.
First, we note that the total spin is conserved, if $\ff B_{r}=0$:
\ba
  &&
  \frac{d}{dt} \left(
  s {\ff n}_{\rm tot} + S \ff m_{\rm tot}
  \right)
  =  
  s \dot{\ff n}_{\rm tot} + \sum_{r} S \dot{\ff m}_{r}
\nonumber \\
  & &
  =
  \frac{\partial H_{\rm eff}(\ff \eta, \ff m)}{\partial \ff \eta} \times \ff \eta
  +
  \sum_{r} \frac{\partial H_{\rm eff}(\ff \eta, \ff m)}{\partial \ff m_{r}} \times \ff m_{r}
  = 0
  \: ,
\nonumber \\
\ea
exploiting \refeq{dheff}.
Hence, $s {\ff n}_{\rm tot} + S \ff m_{\rm tot} = \mbox{const}$.
Energy conservation $(d/dt) H_{\rm eff}(\ff \eta, \ff m) = 0$ follows by construction and can also be verified explicitly.
There are further conserved quantities. 
From \refeq{eomtb} we immediately find $|\ff m_{r}|=\mbox{const.} = 1$, $|\ff \eta|=\mbox{const.} = 1$, and for $\ff B_{r} = 0$ we can derive
\ba
\Delta \dot{\ff \eta} 
&=&
S K \ff m_{0} \times \ff \eta
\; , 
\nonumber \\
\dot{\ff m}_{0}
&=& 
s K \ff \eta \times \ff m_{\rm tot} 
\; ,
\nonumber \\
\dot{\ff m}_{\rm tot}
&=& 
s K \ff \eta \times \ff m_{0} 
\, .
\ea
Further, \refeq{eomtb} yields $\ff m_{0} \ff m_{\rm tot} = \mbox{const.}$, and for $R=2$, in particular, we trivially have $\ff m_{0} \ff m_{\rm tot} = 0$. 
Generally, one {\em cannot} infer $m_{0} = \mbox{const.}$ or $m_{\rm tot} = \mbox{const.}$, opposed to the ASD and \refeq{cons}.
We rather have $\ff m_{0}^{2} + \ff m_{\rm tot}^{2} = \mbox{const.}$ and $(\ff m_{0} \pm \ff m_{\rm tot})^{2}= \mbox{const.}$ only.
We conclude that the effective spin dynamics under tight-binding constraints differs from the naive adiabatic theory as well as from ASD.

\section{Non-admissible effective Lagrangian in the case $\Delta=0$}
\label{sec:delta0}

We proceed with the discussion of the topologically trivial case $\Delta=0$. 
Here, \refeq{eomtb} implies that $\ff m_{0} \times \ff \eta = 0$, and this yields $\ff \eta = \pm \ff m_{0} / m_{0}$, and $\ff \eta = z_{K} \ff m_{0} / m_{0}$ {\em if}, initially, the dynamics starts with the ground-state configuration of the host spins for given $\ff m$. 
We would thus exactly recover the naive adiabatic theory, see Sec.\ \ref{sec:eff}, and \refeq{ndyn} in particular. 
However, the constrained Lagrangian theory is inconsistent in general, as $\ff m_{0} \times \ff \eta = 0$ may conflict with an initial state of the system where $\ff m_{0} \times \ff \eta \ne 0$. 

In the case $\Delta = 0$, one can in fact show that constraining the spin system by imposing \refeq{tcon} leads to a singular effective Lagrangian.
This singularity is subtle. 
For a discussion, we first start with a short note on Lagrange mechanics for a system of point particles described by $N$ coordinates $q=(q_{1},...,q_{N})$. 
Consider the Euler-Lagrange equations
\ba
0 &=& \frac{\partial L(q,\dot{q})}{\partial q_{j}} - \frac{d}{dt}\frac{\partial L(q,\dot{q})} {\partial \dot{q}_{j}} 
\nonumber \\
&=& \frac{\partial L}{\partial q_{j}} 
- \sum_{i} \frac{\partial^{2} L} {\partial {q_{i}} \partial \dot{q}_{j}} \dot{q_{i}}
- \sum_{i} \frac{\partial^{2} L} {\partial \dot{q_{i}} \partial \dot{q}_{j}} \ddot{q_{i}}
\: .
\ea
The Lagrangian is called singular, if the Hesse matrix $H_{ij} = \partial^{2} L / \partial \dot{q_{i}} \partial \dot{q}_{j}$ 
cannot be inverted. 
This is the typical case in classical spin dynamics and explains why there is no simple connection between Hamiltonian and Lagrangian formalism mediated by a Legendre transformation (see, e.g., the discussion in the supplemental material of Ref.\ \cite{EMP20}, section B) and why it is convenient to stay with in Lagrangian framework when discussing constrained classical spin systems.
In principle, however, a Hamiltonian formulation can be derived directly from a singular Lagrangian with the Dirac-Bergmann formalism \cite{Dir51,AB51,BGJN56}.
However, the problem is more severe if $L(q,\dot{q}) = L(q)$. 
In such a case not only the Hesse matrix is singular (in fact, $H=0$), but also the coefficient matrix $K_{ij} = \partial^{2} L / \partial {q_{i}} \partial \dot{q}_{j}$ vanishes.
This may lead to inconsistencies and, hence, such Lagrangians are not admissible.
This means that imposing the constraint is unphysical and does not lead to a valid effective theory. 

The latter exactly applies to our spin system when imposing the constraint (\ref{eq:tcon}) in the case $\Delta=0$.
This can be seen by a gauge transformation of the effective Lagrangian.
We use the constraint \refeq{tcon} explicitly to rewrite the effective Lagrangian (\ref{eq:leff0}). 
With 
\be
\frac{d}{dt} \ff n_{0,i}(\ff \eta) = (\dot{\ff \eta} \nab) \ff n_{0,i}(\ff \eta) = z_{i} \dot{\ff \eta}
\ee
we find 
\ba
L_{\rm eff}(\ff \eta, \dot{\ff \eta}, \ff m,\dot{\ff m}) 
&=&
S \sum_{j} \ff A(\ff m_{j}) \dot{\ff m}_{j}
+
s \sum_{i} \ff A(z_{i} \ff \eta) 
z_{i} \dot{\ff \eta}
\nonumber \\
&-&
H_{\rm eff}(\ff \eta, \ff m) \; .
\labeq{leff2}
\ea
The curl of the vector potential $\ff A(\ff r) \equiv \ff A_{\ff e_{z}}(\ff r) = - (1/r^{2}) (\ff e_{z} \times \ff r) / (1 + \ff e_{z} \ff r / r)$ is invariant under a gauge transformation that replaces $\ff e_{z}$ by an arbitrary unit vector $\ff e$. 
The Euler-Lagrange equations are in fact invariant under a {\em local}, $i$-dependent gauge transformation, specified by $\ff e_{z} \mapsto z_{i} \ff e_{z}$, of the second term on the right-hand side:
\be
  \ff A_{\ff e_{z}}(z_{i} \ff \eta) \mapsto \ff A_{z_{i} \ff e}(z_{i} \ff \eta) 
  = 
  - \frac{1}{\eta^{2}} \, \frac{\ff e_{z} \times \ff \eta} {1 + \ff e_{z} \ff \eta / \eta}
  \: . 
\ee
Note that this does no longer depend on the site index $i$.
The result of the transformation is that the second term on the right-hand side of \refeq{leff2} vanishes if $\Delta = \sum_{i} z_{i}=0$ and, hence, the transformed but equivalent effective Lagrangian,
\be
L_{\rm eff}(\ff \eta, \ff m,\dot{\ff m}) 
=
S \sum_{j} \ff A(\ff m_{j}) \dot{\ff m}_{j}
-
H_{\rm eff}(\ff \eta, \ff m) \; ,
\labeq{leff3}
\ee
lacks the dependence on $\dot{\ff \eta}$.
Therefore, it is not admissible. 

\section{Single impurity spin coupled to a magnetic field}
\label{sec:single}

For $R=1$, i.e., for a single impurity spin $\ff S = S \ff m$, and for $\Delta \ne 0$, \refeq{eomtb} reads:
\be
\Delta \dot{\ff \eta} = SK \ff m \times \ff \eta \; , \quad
\dot{\ff m} = sK \ff \eta \times \ff m - \ff B \times \ff m \: ,
\ee
where we have set, without loss of generality, $z_{i_{1}} = +1$.
This implies
\be
\dot{\ff m} = - \frac{s}{S} \Delta \dot{\ff \eta} - \ff B \times \ff m \: .
\labeq{meq}
\ee
For given $\ff m$, the ground state of the tightly bound host-spin subsystem is given by $\ff n_{i} = z_{i} \ff \eta$ with $\ff \eta = z_{K} \ff m = - \mbox{sign} K \, \ff m$.
Taking the time derivative of this additional condition yields: 
\be
  \dot{\ff \eta} = z_{K} \dot{\ff m} \; .
\labeq{dot}  
\ee
Inserting this relation into \refeq{meq}, we find in case of antiferromagnetic Kondo coupling ($z_{K}<0$), antiferromagnetic host-spin structure and odd $L$ ($\Delta = +1$)
\be
\dot{\ff m} = \frac{1}{1- s/S} \, \ff m \times \ff B \: .
\ee
This describees precession with a renormalized frequency as predicted by the ASD, see Ref.\ \cite{EMP20} and \refeq{anoom}.
It seems that the ASD can be re-derived by imposing the above {\em additional} constraint. 
For the general case of arbitrary $R$, however, we must carefully base the considerations on the action principle, as shown below, since using \refeq{dot} to simplify the equation of motion lacks formal justification.

\section{Alternative derivation of the ASD}
\label{sec:alt}

Interestingly, one can indeed give an alternative derivation of the ASD by starting from the effective spin dynamics under tight-binding constraints discussed above and by imposing the {\em additional} constraint 
\be
  \ff \eta = \ff \eta_{0}(\ff m) \equiv z_{K} \ff m_{0} / m_{0} \: ,
\labeq{addcon}  
\ee  
with $z_{K} = - K/|K|$, i.e., assuming that the mutually bound host spins are, at any instant of time, in the ground-state configuration corresponding to the respective impurity-spin configuration. 

To prove our claim, we start from the effective Hamiltonian \refeq{effham}. 
Inserting the constraint, \refeq{addcon}, in the effective Hamiltonian $H_{\rm eff}(\ff \eta, \ff m)$, yields an effective Hamiltonian depending on the impurity-spin degrees of freedom only,
\be
  H_{\rm eff}(\ff m) 
  =
  E_{0}
  -
  |K| s S m_{0}
  -
  S \sum_{r=1}^{R} \ff m_{r} \ff B_{r} \: ,
\ee
which, of course, equals the one derived earlier, see \refeq{heff}.
For the derivation of the effective equation of motion for $\ff m$ under the additional constraint, we use the action principle and follow the steps outlined in Ref.\ \cite{EMP20}. 
The unconstrained dynamics is governed by the Lagrangian
\ba
L_{\rm eff}(\ff \eta, \dot{\ff \eta} , \ff m, \dot{\ff m})
&=&
\ff A(\ff \eta) s\Delta \dot{\ff \eta}
+
\sum_{r} \ff A(\ff m_{r}) S \dot{\ff m}_{r} 
\nonumber \\
&-&
H_{\rm eff}(\ff \eta, \ff m)
\: .
\ea
Using the constraint \refeq{addcon}, we get an effective Lagrangian depending on $\ff m, \dot{\ff m}$ only:
\ba
L_{\rm eff}(\ff m, \dot{\ff m})
&=&
\ff A(\ff \eta_{0}(\ff m)) \, s \Delta
\sum_{r} (\dot{\ff m}_{r} \nab_{r}) \ff \eta_{0}(\ff m)
\nonumber \\
&+&
\sum_{r} \ff A(\ff m_{r}) S \dot{\ff m}_{r}
-
H_{\rm eff}(\ff m) \; .
\ea
This form of $L_{\rm eff}$ differs only slightly from the one discussed in Ref.\ \cite{EMP20} such that 
the resulting Lagrange equations have exactly the same form as \refeq{asd}:
\be
S \dot{\ff m}_{r} = \frac{\partial H_{\rm eff}(\ff m)}{\partial \ff m_{r}} \times \ff m_{r} + \ff T_{r} \times \ff m_{r} \, .
\ee
Here $\ff T_{r}$ is given via 
\be
  T_{r\mu} =  T_{r\mu}(\ff m,\dot{\ff m}) = \sum_{s\nu} \Omega_{r\mu,s\nu}(\ff m) \dot{m}_{s\nu}  \: ,
\label{eq:tjapp}  
\ee
in terms of 
\be
\Omega_{r\mu,s\nu}(\ff m) = 4\pi s \Delta e_{r\mu,s\nu}(\ff m) \: ,
\labeq{omdefs}
\ee 
which differs from \refeq{omdef} by the missing sum over $i$ and the additional factor $\Delta$.
We have:
\be
e_{r\mu,s\nu}(\ff m)
=
\frac{1}{4\pi}
\frac{\partial \ff \eta_{0}(\ff m) }{\partial m_{r\mu}}
\times
\frac{\partial \ff \eta_{0}(\ff m) }{\partial m_{s\nu}}
\cdot
\ff \eta_{0}(\ff m) 
\: .
\label{eq:cds}
\ee
This topological charge density can be computed as in Sec.\ \ref{sec:top}, and we find
\be
e_{r\mu,s\nu}(\ff m)
=
\frac{1}{4\pi}
z_{K} z_{i_{r}} z_{i_{s}} 
\sum_{\tau}
\varepsilon_{\mu\nu\tau}
m_{0\tau}
\frac{1}{m^{3}_{0}} 
\: .
\ee
Therewith, we have 
\be
  \Omega_{r\mu,s\nu}(\ff m) 
  = 
  z_{i_{r}} z_{i_{s}} z_{K} s \Delta 
  \sum_{\tau} \varepsilon_{\mu\nu\tau} m_{0\tau} \frac{1}{m^{3}_{0}} \: .
\ee  
This is exactly the result found earlier, see \refeq{omres}, and thus yields the same expression for the topological spin torque, \refeq{sumtm}, and the same equations of motion, \refeq{tdyn}, for $\ff m_{r}$.

\end{document}